\documentclass[]{aastex631}

\usepackage{latexsym,amsmath,amssymb}
\usepackage{graphics,graphicx}
\usepackage{natbib}
\usepackage{rotating}
\newcommand{\degmark}{$^\circ$}

\def \rchisq {$\chi_{\nu} ^{2}$}

\newcommand{\asca}{{\it ASCA}}
\newcommand{\rosat}{{\it ROSAT}}

\newcommand{\xmm}{{\it XMM-Newton}}

\newcommand{\swi}{{\it Swift}}
\newcommand{\cha}{{\it Chandra}}
\newcommand{\nustar}{\textit{NuSTAR}}

\newcommand{\msun}{$M_{\odot}$}
\newcommand{\rsun}{$R_{\odot}$}

\newcommand{\lumcgs}{ergs~s$^{-1}$}
\newcommand{\chisq}{$\chi ^{2}$}

\shorttitle{Advective hot flows in BZ Cam and V592 Cas}
\shortauthors{Balman et al.}
\graphicspath{{./}{figures/}}

\begin{document}

\title{Characterizing the Advective Hot Flows of Nova-Like Cataclysmic Variables in the X-rays: The case of BZ Cam and V592 Cas}

\correspondingauthor{\c{S}\"olen Balman}
\email{solen.balman@gmail.com,solen.balman@istanbul.edu.tr}

\author[0000-0001-6135-1144]{\c{S}\"olen Balman}
\affiliation{Department of Astronomy and Space Sciences, Faculty of Science,  Istanbul University, Beyazit, 34119, Istanbul, Turkey}
\affiliation{Kadir Has University, Faculty of Engineering and Natural Sciences, Cibali 34083, Istanbul, Turkey}

\author[0000-0002-4162-8190]{Eric M. Schlegel}
\affiliation{Department of Physics and Astronomy, University of Texas-San Antonio, San Antonio, TX 78249, USA }

\author[0000-0002-4806-5319]{Patrick Godon}
\affiliation{Department of Astrophysics \& Planetary Science, Villanova University, 800 Lancaster Avenue, Villanova, PA 19085, USA}
\affiliation{Henry A. Rowland Department of Physics \& Astronomy, Johns Hopkins University, Baltimore, MD 21218, USA}









\begin{abstract}

We present a joint spectral analysis of  \rosat\ PSPC, \swi\ XRT, and \nustar\ FPMA/B  data of the nova-like (NL) cataclysmic variables (CVs), BZ Cam and V592 Cas in the 0.1-78.0 keV band.  
Plasma models of collisional equilibrium  fail to model the 6.0-7.0 iron line complex and continuum with \rchisq\ larger than 2.0. 
Our results show  nonequilibrium ionization (NEI) conditions in the X-ray plasma with temperatures
of 8.2-9.4 keV and 10.0-12.9 keV for BZ Cam and V592 Cas, respectively.  The centroids of  He-like and H-like iron ionization lines are not at their equilibrium values as expected from NEI conditions. 
We find power law spectral components that reveal the existence of scattering and Comptonization with a photon index of 
1.50-1.87.  We detect a P Cygni profile  in the H-like iron line of BZ Cam translating to outflows of 4500-8700 km s$^{-1}$ consistent with the fast winds 
in the optical and UV. {\it This is the first time such a fast collimated outflow is detected in the X-rays from an accreting CV}.  
An Iron K$\alpha$ line around 6.2-6.5 keV is found revealing the existence of reflection effects in both sources.  
We study the broadband noise  and find that the optically thick disk truncates in BZ Cam and V592 Cas consistent
with transition to an advective hot flow structure. V592 Cas also exhibits a quasi-periodic oscillation  at 1.4$^{+2.6}_{-0.3}$ mHz.
In general, we find that the two NLs portray spectral and noise characteristics as expected from advective hot accretion flows at low radiative efficiency. 

\end{abstract}



\section{Introduction} \label{sec:intro}

Cataclysmic Variables (CVs) are close binary systems where
a white dwarf (WD) accretes matter from a late-type Roche Lobe filling main sequence star
\citep{1995Warner}. In non-magnetic CVs the transferred material
forms an accretion disk that reaches the WD.
Standard accretion disk theory \citep{1973Shakura} predicts half of the accretion
luminosity to emerge
from the disk and the other half from the
boundary layer (BL) very close to the WD \citep{1974Lynden-Bell}. In the standard disk accretion model, 
the BL emission is such that  during low-mass accretion states
it is optically thin emitting in the hard X-rays \citep{1993Narayan,1999Popham} and for high accretion rate states ($\dot M_{acc}$$\ge$10$^{-(9-9.5)}$M$_{\odot}$), it is
optically thick emitting in the soft X-rays and EUV (kT$\sim$10$^{(5-5.6)}$ K; \citep{1995Popham,1995Godon,2014Suleimanov,2015Hertfelder,2017Hertfelder}.
\citet{1993Narayan} show that the optically thin BLs  can be radially extended, advecting part of the
accretion energy to the WD as a result of their inability to cool.
The standard disk is often found inadequate to model disk-dominated, high state CVs (i.e., nova-likes) as well as several quiescent dwarf nova (e.g., eclipsing) in the optical and UV. In high state systems, the standard disk spectrum is bluer than the observed UV spectra indicating that the expected hot optically thick inner flow of the BL is not existent  \citep{2007Puebla,2010Linnell}.  As  a result,  standard disk models  with a truncated inner disks have been calculated  \citep[e.g.,][]{2017Godon}
and have been successfully fitted to the UV spectra while modeling the cooler disk. In addition, an improved model produced for a particular NL fits the FUV spectral slope and its features well, indicating that not much of a vertical structure of temperature exists and most dissipation occurs from the heated surface layers instead of the midplane of the disk \citep{2021Hubeny}. In general, UV analyses find extended emission above the disk supporting nonstandard accretion flows \citep{2007Puebla,2011Puebla}. However, the extent and nature of the X-ray flows are not fully  justified in accordance with these new models.  
It has been suggested that the accretion flows in nonmagnetic white dwarf binaries are well explained in the context of radiatively inefficient advective hot flows in the X-ray regime \citep{2021Mereghetti,2020Balman,2012Balman}. 
This work is extrapolation and development  of our previous works on the two NLs regarding ADAF-like accretion flows in the X-rays \citep{2014Balman,2017Godon}.
Such radiatively inefficient hot flow structure explain most of the complexities in the X-rays and other wavelengths 
as opposed to a standard optically thick accretion flow that is thought to exist in the inner disk of accreting  white dwarf binaries (AWBs) \citep[see][for a review]{2020Balman}.

The nonmagnetic nova-likes (NLs) are found
mostly in a state of high mass accretion rate (a few $\times$10$^{-8}$\msun\ yr$^{-1}$ to a few $\times$10$^{-9}$ \msun\ yr$^{-1}$). The VY Scl-type subclass exhibits
high states and occasional low states of optical brightness while the UX UMa sub-type
remains in the high state \citep{1995Warner}. One of the two NLs discussed in this paper, BZ Cam is a VY Scl-type system whereas the other NL, V592 Cas, is of UX UMa type. 
All NLs show emission lines in the optical and/or UV wavelengths and signatures of winds in these bands.
Bipolar outflows and/or rotationally dominated winds from NLs are
detected typically in the FUV by the P Cygni profiles of the resonance doublet of
CIV \citep{1985Sion}. Mass loss rates of winds  are $\le$ 1\% of the accretion rate,
with velocities of 200-5000 km s$^{-1}$ \citep{2004Kafka,2002Long}.

\subsection{BZ Cam and V592 Cas} \label{sec:NLs}

BZ Cam is an interesting NL classified as an VY Scl system among CVs with a period of 221 min \citep{1996Patterson}. 
It has a bow-shock nebula \citep{1987Krautter,1995Griffith,2001Greiner} which is not consistent with a planetary nebula origin. New findings
indicate that it may be associated with  recurrent nova explosions of BZ Cam in timescales of millenia \citep{2020Hoffmann,2020Tappert}. 
Winds from BZ Cam are detected in the FUV resonance lines (as P Cygni profiles) along with the Balmer and He I lines .
They show a bipolar nature with unsteady and continuously variable outflow of $\sim$3000-5300 km s$^{-1}$ 
and  show time variability from 100s to 3000s in the optical and UV indicating episodic behavior \citep{2013Honeycutt}.
\citet{2001Greiner} present analysis of an optical low state which shows that P Cygni profiles are absent in this state. 
\citet{2017Godon} find a WD temperature of 45000$\pm$5000 K using FUSE and IUE data. The binary system has an inclination 
of about i = 12\degmark-40\degmark\ .  A distance of 830$\pm$160 pc \citep{1998Ringwald} was long accepted, but the 
GAIA archive\footnote{https://gea.esac.esa.int/archive}\  yields a distance of 374$\pm$3 pc using the parallax measurement 
(we will use the GAIA value in this paper). 

V592 Cas is an UX UMa subtype with an inclination of $i=28^{\circ}$$\pm$10$^{\circ}$
\citep{1998Huber}, and a period of 165 min \citep{1998Taylor} with positive and negative superhumps. 
The GAIA distance is 466$\pm$4 pc. The mass of the 
WD in V592 Cas is about 0.75 M$_{\odot}$ with a temperature of 45,000K \citep{2009Hoard}
and a mass accretion rate around 1$\times$10$^{-8}$ M$_{\odot}$ yr$^{-1}$ \citep{1998Taylor,2009Hoard}. 
V592 Cas has a bipolar wind outflow which is episodic and reaches velocities of 5000 km s$^{-1}$ measured in H$\alpha$. Moreover,
the optical brightness variations and the  strength of the outflow are not correlated \citep{2009Kafka}. 
V592 Cas was observed with IUE \citep{1998Taylor} and FUSE \citep{2004Prinja} revealing
blue-shifted absorption troughs in the UV with a non-sinusoidal behavior over the orbital phase. 
The outflowing wind does not show modulation over the negative or positive superhump periods of the source, but is modulated on the orbital period.
Nevertheless, the system indicates existence of a warped, precessing accretion disk . 

\subsection{Previous X-ray observations}\label{int:xray}

At low mass accretion rates, nonmagnetic CV observations
(i.e., dwarf nova  in quiescence)
have yielded  hard X-ray spectra that shows consistency with an optically thin
multi-temperature isobaric cooling flow model of plasma emission (kT$_{max}$$\simeq$10-50 keV) 
\citep[see][and references therein]{2005Pandel,2006Kuulkers,2011Balman,2017Wada,2020Balman}.
At high mass accretion rates ($\dot M_{acc}$$\ge$10$^{-9}$ M$_{\odot}$ yr$^{-1}$),
as opposed to expectations from standard steady-state accretion flow scenarios, observations of NLs
have always shown a hot optically thin X-ray source \citep{1985Patterson,1996vanTeeseling}.  Furthermore, studies with  data from
 \asca, \xmm, \cha\ and \swi\ have been modeled with double MEKAL models or multi-temperature plasma models
with luminosities $\le$ a few $\times$10$^{32}$ \lumcgs\  \citep{2002Mauche,2004Pratt,2014Page,2014Zemko,2014Balman,2017Dobrotka}. 
In the light of these analyses of NLs, the optically thin hard X-ray emission with virial temperatures in the X-ray emitting region have been interpreted as the existence of
ADAF-like radiatively inefficient advective accretion flows (RIAF ADAF)  in the X-ray emitting zones near the WD \citep{2014Balman,2015Balman,2020Balman,2021Kimura}. 
The existence of such RIAF ADAF-like flows, and the transition radii into such nonstandard flows in CVs have also been articulated and  
studied using broadband noise analysis in nonmagnetic CV systems 
(mainly dwarf novae)  by  \citet{2012Balman,2015Balman,2020Balman} .  This aperiodic timing variability is studied and reviewed in \citet{2019Balman,2020Balman} 
where the break frequencies in the characteristic red noise structure of the power spectra show the change  in and the diminishing Keplerian flow in a standard Keplerian disk into a sub-Keplerian RIAF ADAF-like flow. 
The range of break frequencies are 1-6 mHz for quiescent dwarf novae translating to a transition radius of  (3-10)$\times$10$^{9}$cm.

 The \swi\  XRT (X-ray) spectrum of BZ Cam  and  V592 Cas have been found consistent with a multi-temperature plasma
emission model in collisional equilibrium at kT$_{max}$=20-50 keV within a 90\% Confidence Level range,
indicating that the X-ray emitting plasma is virialized  \citep{2014Balman}. \swi\ does not detect any iron emission lines in the 6.0-7.0. keV range.
In this study, 7 eV upper-limit for any blackbody temperature of a soft X-ray component was  also calculated using \rosat\ data of the two sources.
The study  also shows that the  ratio (L$_{x}$/L$_{disk}$) (L$_{disk}$ from the UV-optical wavelengths)
yields considerable inefficiency of emission in the X-ray emitting region with $\sim$ 0.01-0.001. This means that the accretion energy is stored in the flow.
 \citet{2014Balman,2020Balman} discuss that this energy goes into heating the WDs in the NLs or is directed to produce outflows or directed back to the outer disk aiding the production of
 the fast winds in these systems (see the discussion section of \citealt{2014Balman}\ for details).  The WDs in CVs can not be heated more than 15$\%$ via accretional heating and this value is limited to 1$\%$ for the very hot WDs
as in these NLs \citep{2003Godon}. If one scales the WD temperatures in our NLs (similar to most NLs) and the WD temperatures in polar-type magnetic CVs, which do not exhibit disk accretion  in the 3-4 hr orbital period range, one finds 
 that the ratio of the luminosities of the WDs is about 100 yielding a requirement for an inefficiency of 0.01 \citep{2014Balman} in the X-ray emission to produce the excessive heating of WDs.
The power-law index of the temperature distribution  shows departures from the isobaric cooling flow plasma in equilibrium as derived from the fits \citep{2014Balman}.
 In addition, a second component in the \swi\ X-ray spectra of BZ Cam recovered at about $3\sigma$ significance is found consistent 
with a power law emission where as V592 Cas has shown this second  power law component around $2\sigma$.  As a result, the X-ray emitting
regions in BZ Cam  and V592 Cas (and in other NL systems) are interpreted as extended optically thin hard X-ray emitting regions 
of radiatively inefficient (ADAF-like) hot accretion flows and/or constitute X-ray corona regions in the inner disk in a fashion similar to X-ray binaries \citep{2014Balman}. 
Moreover, radiatively inefficient (ADAF-like) accretion flows can aid fast collimated outflows from disks as they
have a positive Bernoulli parameter (the sum of the kinetic energy, potential energy and enthalpy) which can consistently explain the formation of
3000-5300 km s$^{-1}$ winds in BZ Cam and V592 Cas (and perhaps other NLs) that are also variable and episodic in nature.

\section{The Observation and data}\label{sec:obs}

BZ Cam and V592 Cas were observed with \nustar\ on 2017 April 29 and 2016 August 28, respectively.
\nustar\ is the first focusing hard X-ray 
space observatory launched on 2012 June 13 \citep{2013Harrison} which hosts two co-aligned optical crystal detectors 
focused onto two focal planes FPMA and FPMB (Focal Plane Module A/B). It covers the energy range 3-79 keV with an effective area summed over the two modules of about 900 cm$^2$ around 10 keV. 
It has an energy resolution of 400 eV at 10 keV and 900 eV at 68 keV with an angular resolution of 18$^{\prime\prime}$ (FWHM). 
For BZ Cam, the source exposure times were 37.8 ks and 37.7 ks for FPMA and FPMB, respectively (ID=30201010002,PI=Balman).
FPMA and FPMB exposures were 39.5 and 39.4 for V592 Cas (ID=30201012002,PI=Balman). 
The data were processed using  NuSTARDAS version 1.7.1, 1.8.0 and  CALDB 20170727. 
The $nupipeline$ tool was run to generate cleaned event lists and remove
passages from South Atlantic Anomaly.   The $nuproducts$ tool was used to extract background and source spectra and light curves together with the appropriate response (RMF) and ancillary (ARF) files.
To extract photons for source and background devoid of contaminations, circular regions with radii 73$^{\prime\prime}$-135$^{\prime\prime}$ for BZ Cam and 50$^{\prime\prime}$-100$^{\prime\prime}$ for V592 Cas were used. The background counts were extracted from different parts of the detector and the extraction sizes were studied and optimized for a 
proper source and background extraction  (as to maximize count rate and statistical quality). We assumed the largest extraction radii in these ranges for the fitted spectra. There were no stray light contaminations associated with the observations. 
Further data reductions were performed using  HEASoft\footnote{https://heasarc.gsfc.nasa.gov/docs/software/heasoft/}\ (version 6.20-6.24). 
In order to increase the S/N in the spectrum, the FPMA/B  
spectra were combined using  HEASoft task  $addspec$ which produced proper response and ancillary files. This step was inspected
carefully as to how well the FPMA/B spectra were incorporated together using the original spectral files of FPMA/B and comparative analysis (using proper response and ancillary files).  The net source count 
rates were 0.113(2) c s$^{-1}$ and 0.071(2) c s$^{-1}$ for BZ Cam and V592 Cas, respectively.
 
As second and third observations, we used the \swi\ XRT and the \rosat\ data of BZ Cam and V592 Cas, that were published in \citet{2014Balman} in detail, to cover the
soft energy ranges and calibrate the spectral fits better.  \swi\ XRT is a focusing X-ray telescope with 110 cm$^2$ effective area at 1.5 keV and a 
24$^{\prime}$ field of view in the 0.3-10.0 keV energy range \citep{2005Burrows}. The data were obtained in the PC (photon couting mode)
with full imaging capacity. The source exposures were 15 ksec and the net
count rates were 0.069(3) c s$^{-1}$ and 0.051(2) c s$^{-1}$, for BZ Cam and V592 Cas, respectively. The \rosat\ count rates were similar with
 0.078(4) c s$^{-1}$ (BZ Cam) and 0.048(8) c s$^{-1}$ (V5923 Cas) in the 0.1-2.0 keV band. BZ Cam observation (ID=rp300233n00) was performed on 1992 September 28-29 (not a low state) using the PSPCB detector (6.0 ksec). Data of V592 Cas (ID=rs930701n00) were obtained during the RASS (\rosat\ all sky survey; 1990 December 29 to 1991 August 06) 
using the PSPCC detector (0.6 ksec).  The spectra for \swi\ and \rosat\ were extracted and analyzed using  similar versions of  HEASoft and XSELECT\footnote{https://heasarc.gsfc.nasa.gov/ftools/xselect/} v2.4-e.

\section{Spectral Analysis}\label{sec:spec}

As described in the previous section, background subtracted spectra, response and ancillary files were generated for BZ Cam and V592 Cas  using  the $nuproducts$ tool for the \nustar\ data. Subsequently, the FPMA/B spectra were combined.
The combined spectra were grouped to yield a S/N ratio of 10-12 in each spectral bin to acquire good \chisq\ statistics.
The soft energies not covered by \nustar\  were supplemented with the \rosat\ PSPC and \swi\ XRT data published and/or used in \citet{2014Balman}. The spectral extraction of the \swi\ XRT 
spectra  are summarized in Sec. 6 of this paper.  Spectra grouped to have a minimum of 30 counts in each bin, are used in this paper 
whereas the fits in  \citet{2014Balman} were performed on spectra with 60 counts in each bin. This is done to utilize the 
spectral resolution of the data for a better match 
to  \nustar\ data. \rosat\ data are used to accommodate for the softest X-ray energies in conjunction with the other two observatory data.  A grouping of
10-20 counts in each bin is used  to achieve adequate statistics. 
The spectral analysis of the joint spectra are done within XSPEC software (for references and model descriptions see \citealt{1996Arnaud}\footnote{https://heasarc.gsfc.nasa.gov/xanadu/xspec/manual/Models.html}). 
To account for the interstellar absorption in the broadband X-ray spectra, the $tbabs$ model was utilized in the fitting procedures (abundances set to "wilm" \citealt{2000Wilms}).  
During the entire spectral analysis, constant multiplicative model factors have been incorporated in the fits to account for the cross-normalization calibration between different observatory data.
We caution that we do not expect a change in long time variability characteristics of our sources (i.e., high or low states) as mentioned in the introduction and Sec.~\ref{sec:NLs}. Moreover, we do not find any sporadic events or flare-type events in any of the observations included from the three missions used in this study. 

As the first step,  joint X-ray spectra (\rosat$+$\swi$+$\nustar) of BZ Cam and V592 Cas are modeled with the multi-temperature isobaric cooling flow type of plasma emission model CEVMKL in XSPEC \citep{1995Liedahl,1996Singh}\ as was done for the \swi\ analysis (i.e., $tbabs\times$CEVMKL; \citealt{2014Balman}).  We assumed "switch set to 2" where the CEVMKL model uses the APEC model for interpolation with ATOMDB\footnote{http://atomdb.org}\ database.  
The fits with the CEVMKL model are conducted between 0.2-75.0 keV yielding unacceptable values above \rchisq  $\sim$ 2.0 with large sigma deviations around the iron line complex between 6.0-7.0 keV for both sources.  The spectral parameters of the fit are kT$_{max}$=28-38 keV and $\alpha <$ 0.02 for V592 Cas at 90\% Confidence Level ($\alpha$ is the power law index parameter for the temperature distribution which is expected to be 1.0 for a cooling flow type of plasma). For BZ Cam, these parameters are  
kT$_{max}$=14-19 keV and $\alpha <$ 0.04 at the same Confidence Level. In the fitting procedure, we assumed fixed solar abundances since the fits  were not sensitive to changing abundances (e.g., iron) where the error ranges include solar abundances.

The low $\alpha$ values are nonphysical for CEVMKL model indicating  significant  lack of radiative cooling in the plasma meaning changes in differential emission measures ($\propto$ L) are very small with changing temperature.
The continuous temperature distribution in the X-rays described with the CEVMKL model is an  isobaric cooling flow type plasma model in collisional equilibrium with a differential emission measure that depends on  a power-law distribution of temperatures ($dEM=(T/T_{max})^{\alpha -1}  dT/T_{max}$). In such a model, the emission measure at each temperature is proportional to the time the cooling gas remains at this temperature \citep{2005Pandel}.
The fitted spectra are displayed in Figure 1. Our analysis shows that
the  $tbabs$$\times$CEVMKL model fits are inadequate. Moreover, such large deviations
in the residuals around 6.0-7.0 keV complex are indicative of nonequilibrium ionization conditions in the plasma flow, since CEVMKL is a collisional ionization equilibrium plasma emission (CIE) model. 
As a result, the CEVMKL model is not used for further analysis since it does not fit the continuum and the iron line complex successfully.  Next, we inspected the behavior of residuals around the iron line complex indicative of other emission and absorption features and  added one gaussian absorption line, using GABS model, and two to three emission lines, using GAUSS model, to account for the variations in the residuals. 
Along these line models, we assumed a Bremsstrahlung continuum model (free-free emission), BREMSS (in XSPEC) and a nonequilibrium ionization plasma emission model, VNEI (in XSPEC). The \swi\ spectra are also fitted better with the addition of neon or iron emission and absorption line features represented with other GABS and GAUSS models in the fitting procedure.
The composite model fits to the joint \rosat, \swi\  and \nustar\ spectra are displayed  in Figure 2 and 3  and results of the spectral fits
are given in Tables 1 and 2, for BZ Cam and V592 Cas, respectively.

\begin{deluxetable*}{ccccc}
\tablewidth{0pt}
\tablecaption{Spectral Parameters of the
Fits to the joint \rosat, \nustar\ and \swi\ Spectra of BZ Cam\label{tab:sp1}}
\tablehead{
Model  & Parameter & Fit-1$^{\S{1}}$ & Fit-1$^{\S{2}}$ &  Fit-2$^{\S{3}}$ 
}
\startdata
TBabs  & $N_H$ & 0.06$^{+0.03}_{-0.01}$  & 0.09$^{+0.03}_{-0.03}$ & 0.08$^{+0.03}_{-0.02}$  \\
       &     (10$^{22}$ cm$^{-2}$) &  &  &   \\
\hline
VNEI &  kT\ (keV)    & $ 8.8^{+0.6}_{-0.6} $ & $ 5.9^{+0.6}_{-0.5} $ &  N/A  \\
      &  $\tau\ (s\ cm^{-3}) $ & $3.9^{+2.0}_{-1.0}\times 10^{11}$ & $3.8^{+5.2}_{-1.2}\times 10^{11}$ &  N/A   \\
      & $K_{VNEI}$ &  $2.1^{+0.1}_{-0.2}\times 10^{-3}$ & $2.0^{+0.3}_{-0.2}\times 10^{-3}$ & N/A    \\
\hline
 BREMSS    & kT\ (keV)       &  N/A &   &  $ 9.7^{+0.7}_{-0.6} $   \\
    & $K_{BREMSS}$ &  N/A & N/A  & $6.3^{+0.4}_{-0.3}\times 10^{-4}$   \\
\hline
Power law & Photon Index &  N/A  &  $1.81^{+0.06}_{-0.05}$ & N/A \\
         & $K_{power law}$ &    N/A  &  $2.4^{+0.4}_{-0.4}\times 10^{-4}$ & N/A \\
\hline
GABS & LineE (keV) & 1.15 (fixed)  & 1.15 (fixed)   & 1.22$^{+0.16}_{-0.17}$ \\
     & $\sigma$ (keV) & 1.20$^{+0.2}_{-0.2}$    & 1.5$^{+0.2}_{-0.1}$   & 0.5 (fixed) \\
     & depth &  1.6$^{+0.3}_{-0.3}$    &  3.3$^{+0.5}_{-0.4}$  &  0.5$^{+0.2}_{-0.1}$ \\
GABS & LineE (keV) & 7.05$^{+0.10}_{-0.05}$    &  7.05$^{+0.08}_{-0.08}$   & 7.02$^{+0.01}_{-0.01}$ \\
     & $\sigma$ (keV) & 0.001$^{+0.013}_{>}$   & 0.001$^{+0.013}_{>}$ & 0.06$^{+0.04}_{-0.03}$ \\
     & depth & 1.0$^{+2.0}_{>}$    & 0.04$^{+1.1}_{>}$ &  1.0$^{+0.15}_{-0.13}$ \\
GAUSS & LineE (keV) &  1.01$^{+0.02}_{-0.03}$     &  1.01$^{+0.02}_{-0.03}$ & 1.01$^{+0.02}_{-0.02}$ \\
     & $\sigma$ (keV) &  0.001$^{+0.06}_{>}$ & 0.001$^{+0.05}_{>}$  &  0.001$^{+0.07}_{>}$ \\
     & $K_{Gauss}$ & $6.7^{+2.3}_{-2.4}\times 10^{-5}$     &  $9.0^{+4.0}_{-3.0}\times 10^{-5}$  &  $6.6^{+2.0}_{-2.0}\times 10^{-5}$  \\
GAUSS & LineE (keV) & 6.38$^{+0.06}_{-0.08}$  & 6.46$^{+0.05}_{-0.07}$  &  6.38$^{<}_{-0.2}$  \\
     & $\sigma$ (keV)   & 0.3$^{+0.07}_{-0.07}$   & 0.3$^{+0.1}_{-0.1}$  &  1.0$^{<}_{-0.2}$\\
     & $K_{Gauss}$ &  1.9$^{+0.4}_{-0.3}\times 10^{-5}$  & 2.2$^{+0.4}_{-0.3}\times 10^{-5}$   &  2.9$^{+0.8}_{-0.7}\times 10^{-5}$ \\
GAUSS & LineE (keV) & N/A   &  N/A & 6.89$^{+0.02}_{-0.01}$     \\
     & $\sigma$ (keV)     &  N/A & N/A & 0.01$^{+0.05}_{>}$   \\
     & $K_{Gauss}$ & N/A   & N/A &  2.6$^{+0.5}_{-0.4}\times 10^{-5}$   \\
GAUSS & LineE (keV) & N/A   & N/A  &  6.55$^{+0.02}_{-0.03}$    \\
     & $\sigma$ (keV)  &  N/A & N/A  & 0.01$^{+0.09}_{>}$  \\
     & $K_{Gauss}$ & N/A & N/A  &   1.6$^{+0.3}_{-0.2}\times 10^{-5}$   \\
\hline
 & $\chi^2_{\nu} (\nu)$  & 1.37 (93)   & 1.28 (92)  & 1.20 (90)   \\
 \hline
 Flux & (10$^{-12}$)  & 5.5$^{+0.5}_{-0.7}$ &  4.7$^{+0.7}_{-0.3}$ &   5.1$^{+1.1}_{-0.6}$     \\
    (thermal)   &   erg~cm$^{-2}$s$^{-1}$  &  &  &     \\
 Luminosity & (ther.) &   (10$^{32}$erg~s$^{-1}$)  &  0.74-1.02 &   \\
 \hline
 Flux & (10$^{-12}$)  & N/A  &   2.9$^{+0.3}_{-0.2}$    &    N/A    \\
    (nonther.)   &   erg~cm$^{-2}$s$^{-1}$  &  &   &     \\
 Luminosity &  (nonther.) &  (10$^{32}$erg~s$^{-1}$)  &  0.46-0.54  &   \\
 \hline
\enddata   
\vspace{0.2cm}
\tablecomments{
{\bf \S{1}}-\footnotesize{$tbabs$$\times$GABS$\times$GABS(VNEI+GAUSS)}; \\
{\bf \S{2}}-\footnotesize{$tbabs$$\times$GABS$\times$GABS(VNEI+power+GAUSS)};\\
{\bf \S{3}}-\footnotesize{$tbabs$$\times$GABS$\times$GABS(BREMSS$\texttt{+}$GAUSS$\texttt{+}$GAUSS$\texttt{+}$GAUSS$\texttt{+}$GAUSS)};\\
{\tt $tbabs$}--(abund=wilm)\citet{2000Wilms}. Solar abundances are assumed in the plasma models when necessary.
Fits are performed using \rosat, \swi\ and \nustar\ data in the 0.2-75.0 keV range. NEIvers.3.0.9 plasma code with ATOMDB database was assumed for VNEI fits.
$N_H$ is the absorbing column, $\tau$ is the ionization timescale (n$_e$t),
$K_{Gauss}$, $K_{VNEI}$, and
$K_{BREMSS}$ are the normalization for the Gaussian line, VNEI and BREMSS models.
The normalization constant of the VNEI, BREMSS models are in
K=(10$^{-14}$/4$\pi$D$^2$)$\times$EM where EM (emission measure) =${\rm \int n_e\ n_H\ dV}$
(integration is over the emitting volume V).
All errors are calculated
at 90\% confidence level for a single parameter.
The unabsorbed X-ray flux and the luminosities are given in the range 0.1-100.0 keV.
For luminosities, the distance of 374 pc is assumed (see Section~\ref{sec:intro}).}
\end{deluxetable*}

\begin{deluxetable*}{cccccc}
\tablewidth{0pt}
\tablecaption{Spectral Parameters of the
Fits to the joint \rosat, \nustar, and \swi\ Spectra of V592 Cas\label{tab:sp2}}
\tablehead{
Model  & Parameter & Fit-1$^{\S{1}}$ & Fit-1$^{\S{2}}$ & Fit-2$^{\S{3}}$ & Fit-2$^{\S{4}}$
}
\startdata
TBabs  & $N_H$ &  0.16$^{+0.09}_{-0.07}$  &  0.40$^{+0.06}_{-0.13}$ &  0.21$^{+0.08}_{-0.08}$  & 0.31$^{+0.10}_{-0.09}$  \\
       &     (10$^{22}$ cm$^{-2}$) &  &   &   &  \\
\hline
VNEI &  kT\ (keV)    & 11.4$^{+1.5}_{-1.3} $ & 4.9$^{+1.2}_{-1.4} $ & N/A   & N/A    \\
      &  $\tau\ (s\ cm^{-3}) $ & 2.2$^{+1.0}_{-0.5}\times 10^{11}$ &   $3.8^{<}_{-1.2}\times 10^{11}$ &  N/A & N/A  \\
      & $K_{VNEI}$ &  $1.3^{+0.1}_{-0.1}\times 10^{-3}$ &   $6.2^{+2.7}_{-1.0}\times 10^{-3}$    &  N/A   & N/A  \\
\hline
 BREMSS    & kT\ (keV)    &  N/A &   N/A  &    $10.9^{+1.4}_{-1.1} $  &  $5.0^{+1.1}_{-1.0} $ \\
    & $K_{BREMSS}$ &  N/A & N/A &  $5.0^{+0.3}_{-0.4}\times 10^{-4}$  & $3.7^{+0.4}_{-0.4}\times 10^{-4}$   \\
\hline
Power law & Photon Index &  N/A &  $1.74^{+0.06}_{-0.06}$ & N/A  & $1.56^{+0.05}_{-0.06}$ \\
         & $K_{power law}$ &  N/A   &  $2.3^{+0.3}_{-0.3}\times 10^{-4}$ & N/A  &   $1.4^{+0.2}_{-0.3}\times 10^{-4}$ \\
\hline
GABS & LineE (keV) & 1.16$^{+0.03}_{-0.02}$  & 1.2$^{+0.05}_{-0.05}$ & 1.18$^{+0.07}_{-0.04}$ & 1.22$^{+0.04}_{-0.06}$ \\
     & $\sigma$ (keV) & 0.01 (fixed)  &  0.01 (fixed) & 0.01 (fixed) & 0.01 (fixed)   \\
     & depth &  6721.7$^{<}_{-6714.7}$    & 107.6$^{<}_{107.2}$ &  39.6$^{<}_{-26.4}$ & 14.2$^{<}_{-14.1}$  \\
GABS & LineE (keV) & 6.9$^{+0.1}_{-0.1}$   & 7.09$^{+0.25}_{-0.25}$ &  6.98$^{+0.05}_{-0.05}$ &  7.00$^{+0.04}_{-0.06}$ \\
     & $\sigma$ (keV) & 0.7$^{+0.1}_{-0.1}$ & 0.2$^{+0.2}_{-0.1}$  &  0.16$^{+0.06}_{-0.08}$  &   0.12$^{+0.1}_{-0.05}$ \\
     & depth & 1.3$^{+0.2}_{-0.2}$    & 2.80$^{<}_{-2.78}$  &   0.4$^{+0.2}_{-0.1}$    & 0.34$^{+0.11}_{-0.09}$   \\
GAUSS & LineE (keV) &  1.08$^{+0.02}_{-0.04}$  &  1.06$^{+0.05}_{-0.06}$  &  1.07$^{+0.04}_{-0.03}$  & 1.06$^{+0.05}_{-0.04}$   \\
     & $\sigma$ (keV) & 0.01 (fixed) & 0.01 (fixed)  & 0.01 (fixed)  & 0.01 (fixed)  \\
     & $K_{Gauss}$ & $2.8^{+1.4}_{-1.4}\times 10^{-5}$   &  $2.8^{+1.6}_{-2.3}\times 10^{-5}$ &  $3.0^{+1.5}_{-1.5}\times 10^{-5}$ &  $2.8^{+1.9}_{-1.8}\times 10^{-5}$    \\
GAUSS & LineE (keV) & 6.5$^{+0.1}_{-0.10}$ & 6.38$^{+0.09}_{-0.23}$  & 6.4$^{+0.2}_{-0.2}$  & 6.36$^{+0.18}_{-0.19}$  \\
     & $\sigma$ (keV)   & 0.5$^{+0.1}_{-0.1}$  &  0.4$^{+0.2}_{-0.2}$ & 0.4$^{+0.2}_{-0.1}$  & 0.4$^{+0.2}_{-0.2}$ \\
     & $K_{Gauss}$ &  3.4$^{+0.6}_{-0.6}\times 10^{-5}$   &    8.0$^{+1.2}_{-3.5}\times 10^{-5}$    & 9.7$^{+3.6}_{-3.4}\times 10^{-6}$  & 7.2$^{+2.6}_{-2.5}\times 10^{-6}$   \\
GAUSS & LineE (keV) & N/A  & N/A  &  6.88$^{+0.11}_{-0.04}$  & 6.89$^{+0.08}_{-0.05}$  \\
     & $\sigma$ (keV)     &  N/A & N/A  & 0.001 (fixed)  & 0.001 (fixed) \\
     & $K_{Gauss}$ & N/A & N/A  &  9.1$^{+4.1}_{-3.8}\times 10^{-6}$  & 5.9$^{+2.9}_{-2.6}\times 10^{-6}$  \\
GAUSS & LineE (keV) & N/A  &  N/A  & 6.64$^{+0.02}_{-0.08}$   & 6.63$^{+0.08}_{-0.07}$   \\
     & $\sigma$ (keV)  &  N/A & N/A &  0.001 (fixed) &  0.001 (fixed) \\
     & $K_{Gauss}$ & N/A & N/A  & 5.4$^{+2.2}_{-1.9}\times 10^{-6}$  & 4.0$^{+1.6}_{-1.4}\times 10^{-5}$  \\
\hline
 & $\chi^2_{\nu} (\nu)$        &  1.71 (109) &  1.60 (107) &  1.76 (106) &  1.64 (104)   \\
\hline
 Flux & (10$^{-12}$)  & 3.9$^{+0.5}_{-0.2}$ &  1.6$^{+0.3}_{-0.3}$ &   4.1$^{+0.6}_{-0.6}$ & 2.1$^{+0.6}_{-0.3}$     \\
    (thermal)   &   erg~cm$^{-2}$s$^{-1}$  &  &  &   &    \\
 Luminosity & (ther.) &   (10$^{32}$erg~s$^{-1}$)  &  0.92-1.23  (no pow) & 0.34-0.71 (with pow) &   \\
 \hline
 Flux & (10$^{-12}$)  & N/A  &   4.0$^{+0.2}_{-0.1}$    &    N/A   &  3.6$^{+0.3}_{-0.2}$    \\
    (nonther.)   &   erg~cm$^{-2}$s$^{-1}$  &  &   &     &  \\
 Luminosity &  (nonther.) &  (10$^{32}$erg~s$^{-1}$)  & &  0.89-1.1  &   \\
\hline
\enddata  
\vspace{0.2cm}
\tablecomments{
{\bf \S{1}}-\footnotesize{$tbabs$$\times$GABS$\times$GABS(VNEI+GAUSS)};
{\bf \S{2}}-\footnotesize{$tbabs$$\times$GABS$\times$GABS(VNEI+power+GAUSS)};\\
{\bf \S{3}}-\footnotesize{$tbabs$$\times$GABS$\times$GABS(BREMSS$\texttt{+}$GAUSS$\texttt{+}$GAUSS$\texttt{+}$GAUSS$\texttt{+}$GAUSS)};\\
{\bf \S{4}}-\footnotesize{$tbabs$$\times$GABS$\times$GABS(BREMSS$\texttt{+}$power$\texttt{+}$GAUSS$\texttt{+}$GAUSS$\texttt{+}$GAUSS$\texttt{+}$GAUSS)};
{\tt $tbabs$}--(abund=wilm)\citet{2000Wilms}. Solar abundances are assumed in the plasma models when necessary.
Fits are performed using \rosat, \swi\ and \nustar\ data in the 0.2-75.0 keV range.  NEIvers.3.0.9 plasma code with ATOMDB database was assumed for VNEI fits.
$N_H$ is the absorbing column, $\tau$ is the ionization timescale (n$_e$t),
$K_{Gauss}$, $K_{VNEI}$, and
$K_{BREMSS}$ are the normalization for the Gaussian line, VNEI and BREMSS models.
The normalization constant of the VNEI, BREMSS models are in
K=(10$^{-14}$/4$\pi$D$^2$)$\times$EM where EM (emission measure) =${\rm \int n_e\ n_H\ dV}$
(integration is over the emitting volume V).
All errors are calculated
at 90\% confidence level for a single parameter. The unabsorbed X-ray flux and the luminosities are given in the range 0.1-100.0 keV.
For luminosities, the distance of 466 pc is assumed (see Section~\ref{sec:intro}).}
\end{deluxetable*}

\begin{figure*}
\begin{tabular}{ll}
\centerline{
\includegraphics[width=3.15in,height=2.5in,angle=0]{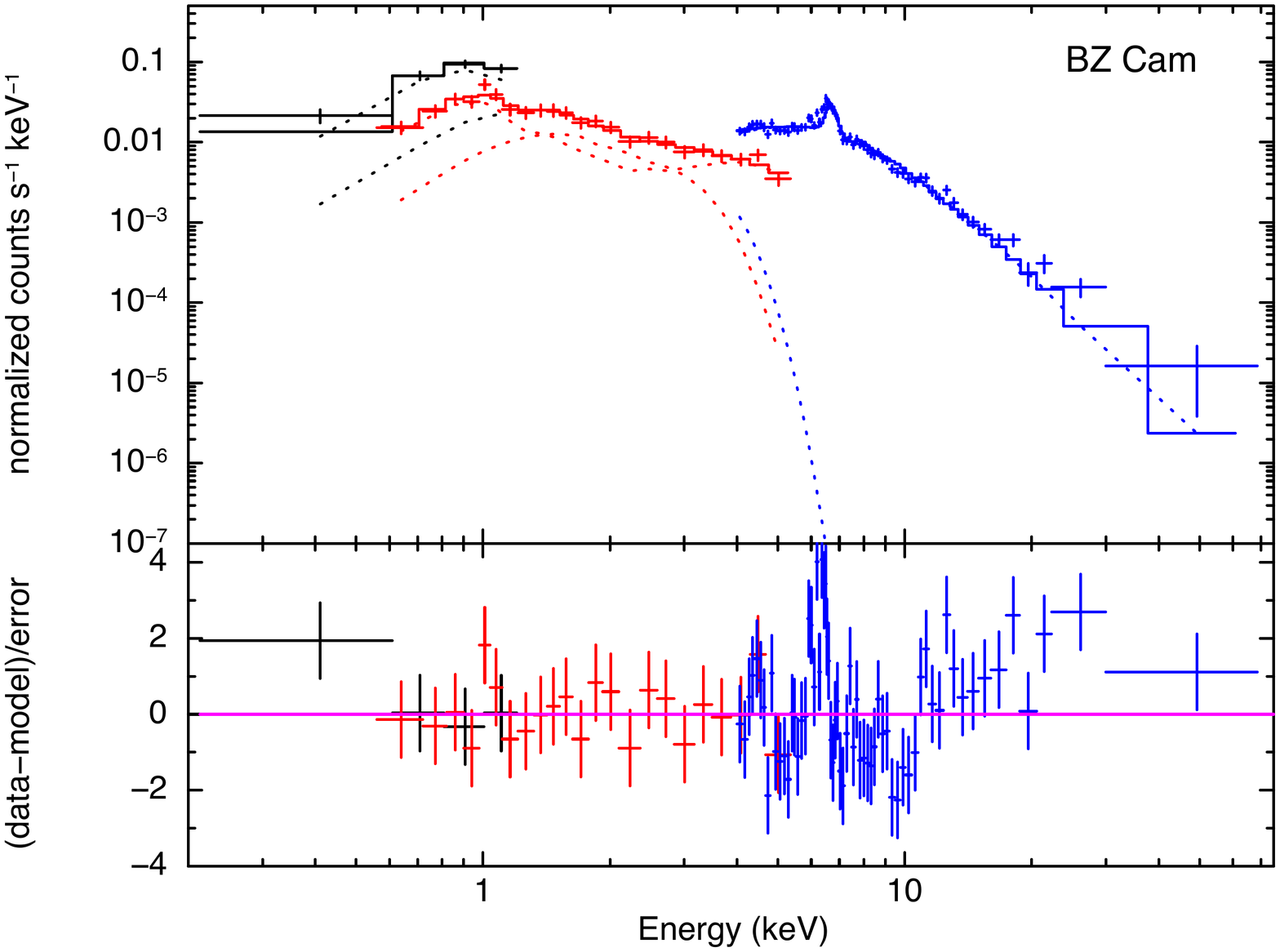} 
\includegraphics[width=3.15in,height=2.5in,angle=0]{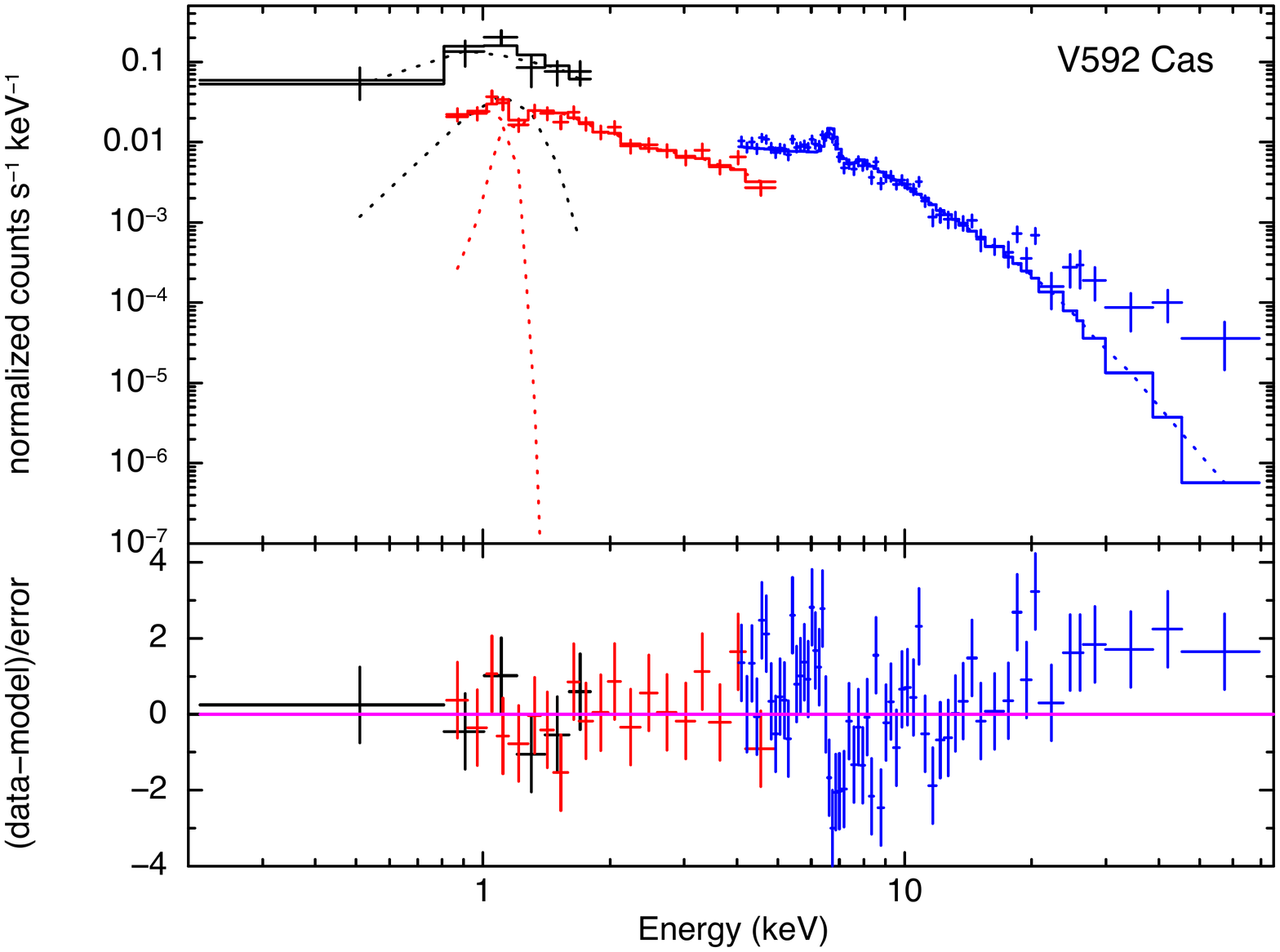}}
\end{tabular}
\caption{The left hand panel is the fitted \rosat, \swi,  and  \nustar\ (combined) spectra of BZ Cam using 
($tbabs \times$CEVMKL) model. The right hand panel is the fitted V592 Cas spectra.
The dotted lines show the individual model components, the crosses are the data (black for \rosat, red for \swi, and blue for \nustar) 
and the solid lines indicate the fitted spectrum. The lower
panels show the residuals in standard deviations.\label{fig1:sp}}
\end{figure*}

\begin{figure*}
\begin{tabular}{lll}
\hspace{-1.0cm}
\includegraphics[width=2.5in,height=2.5in,angle=0]{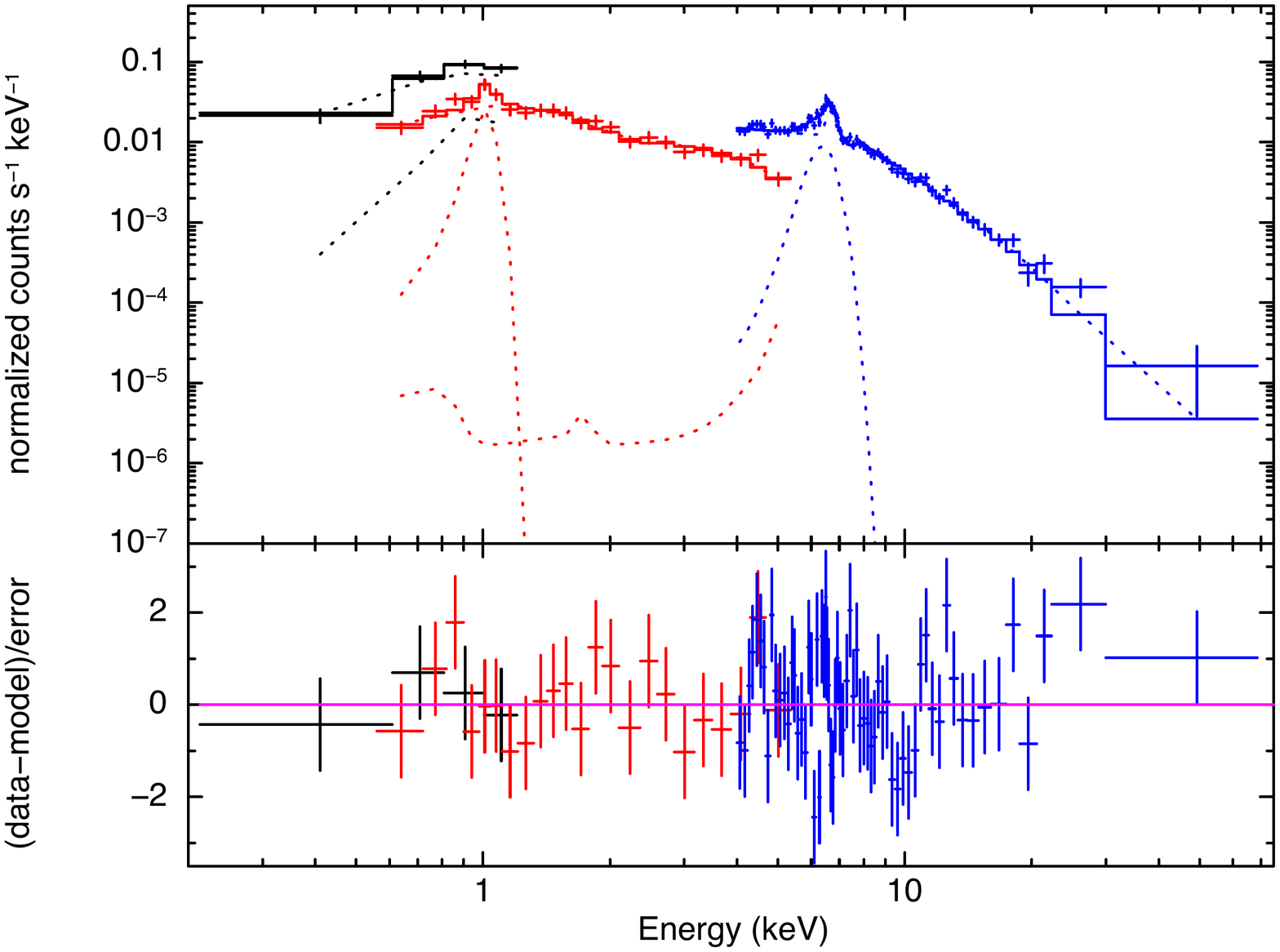} &
\hspace{-0.8cm}\includegraphics[width=2.5in,height=2.5in,angle=0]{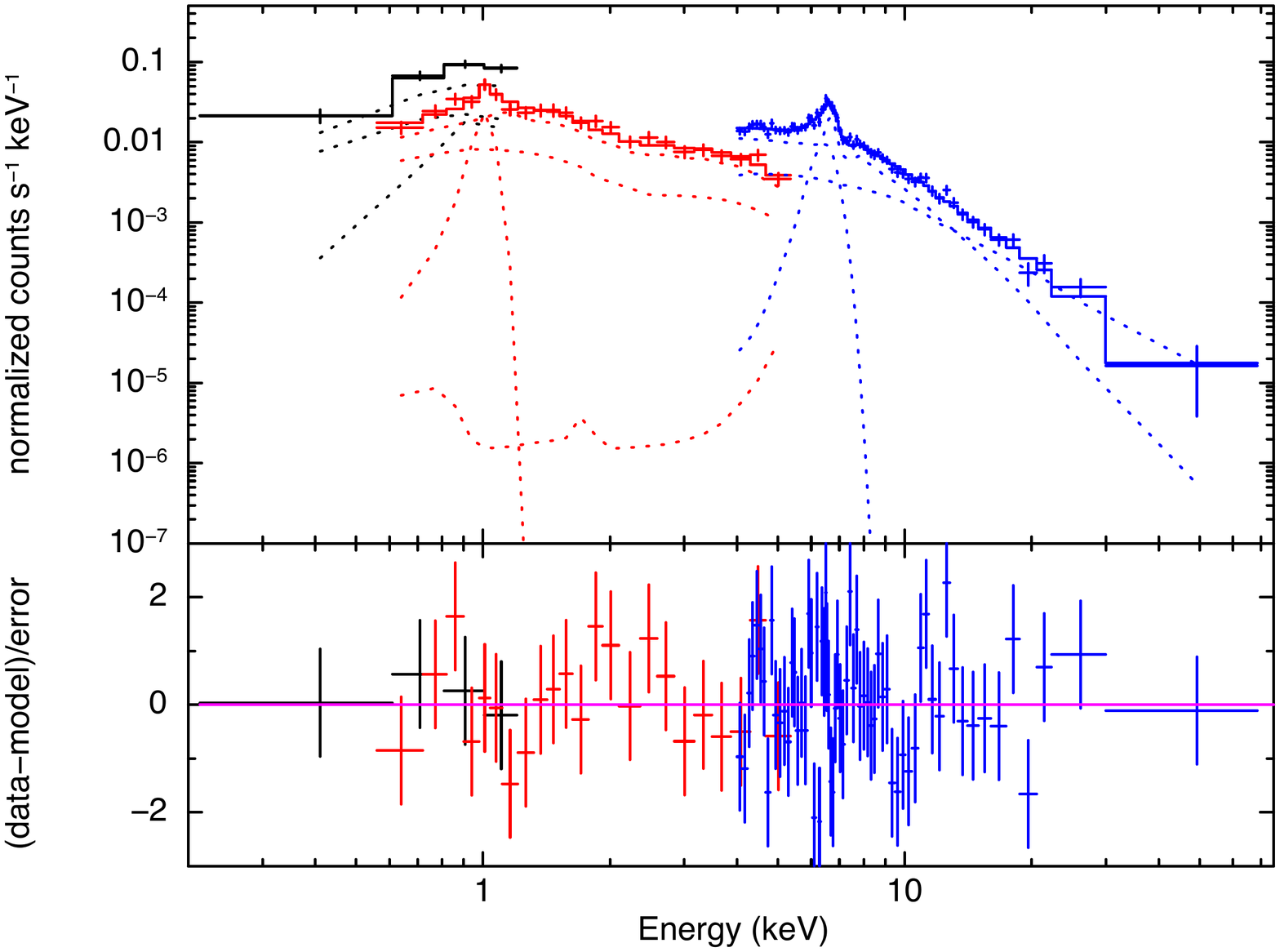} &
\hspace{-0.8cm}\includegraphics[width=2.5in,height=2.5in,angle=0]{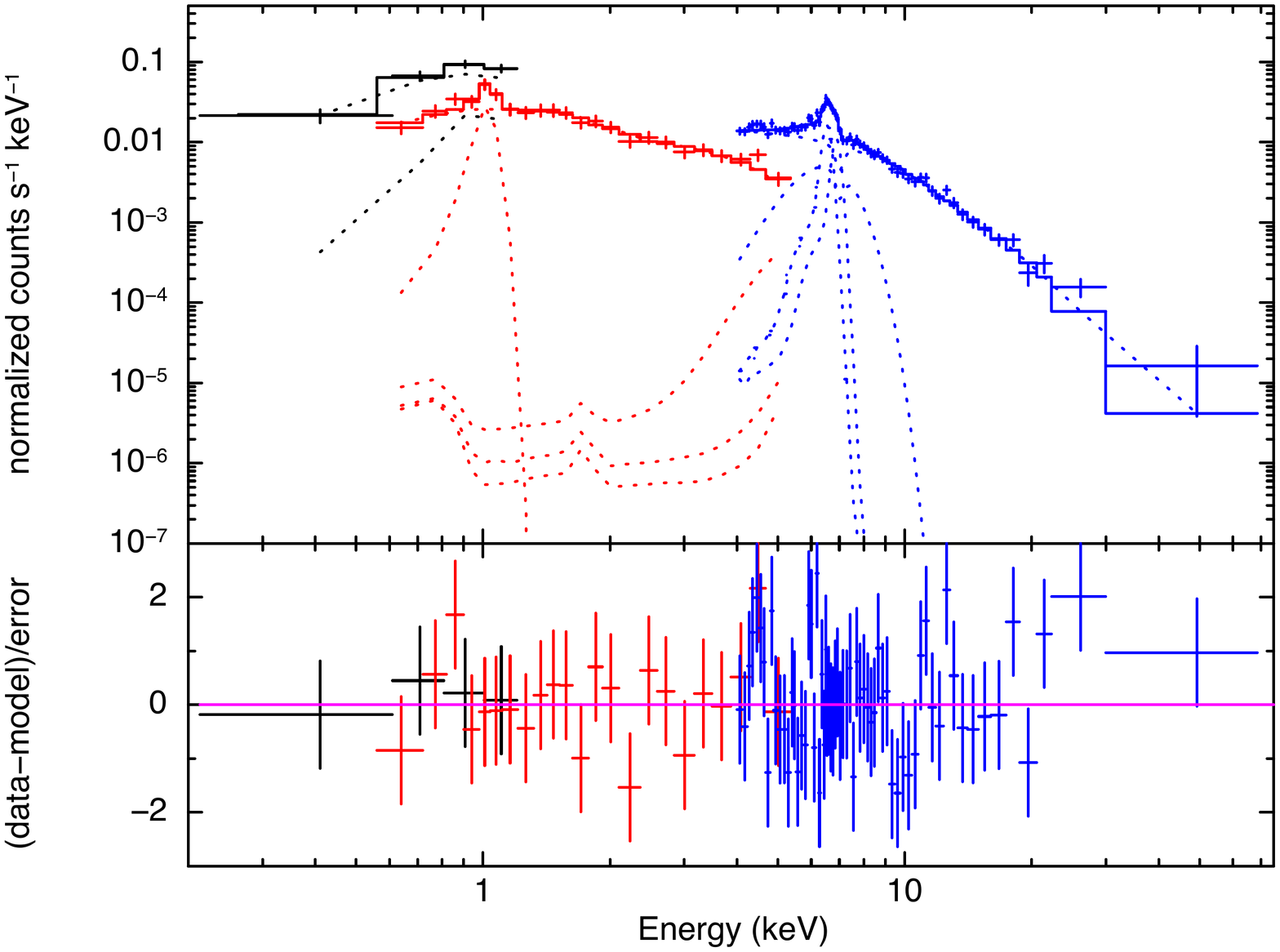}
\end{tabular}
\caption{The left hand panel is the fitted \rosat, \swi, and  \nustar\ (joint) spectra of BZ Cam using the model,
$tbabs$$\times$GABS$\times$GABS(VNEI+GAUSS). The middle panel is the joint spectra fitted with the composite model 
$tbabs$$\times$GABS$\times$GABS(VNEI+power+GAUSS) and the right hand panel shows the same spectra fitted with the model 
$tbabs$$\times$GABS$\times$GABS(BREMS$\texttt{+}$GAUS$\texttt{+}$GAUS$\texttt{+}$GAUS$\texttt{+}$GAUS) as listed in Table  1. 
The dotted lines show the individual
model components, the crosses are the data (black for \rosat, red for \swi, and blue for \nustar) 
and the solid lines indicate the fitted spectrum. The lower
panels show the residuals in standard deviations.\label{fig2:sp}}
\end{figure*}

\begin{figure*}
\begin{tabular}{lll}
\hspace{-1.0cm}
\includegraphics[width=2.5in,height=2.5in,angle=0]{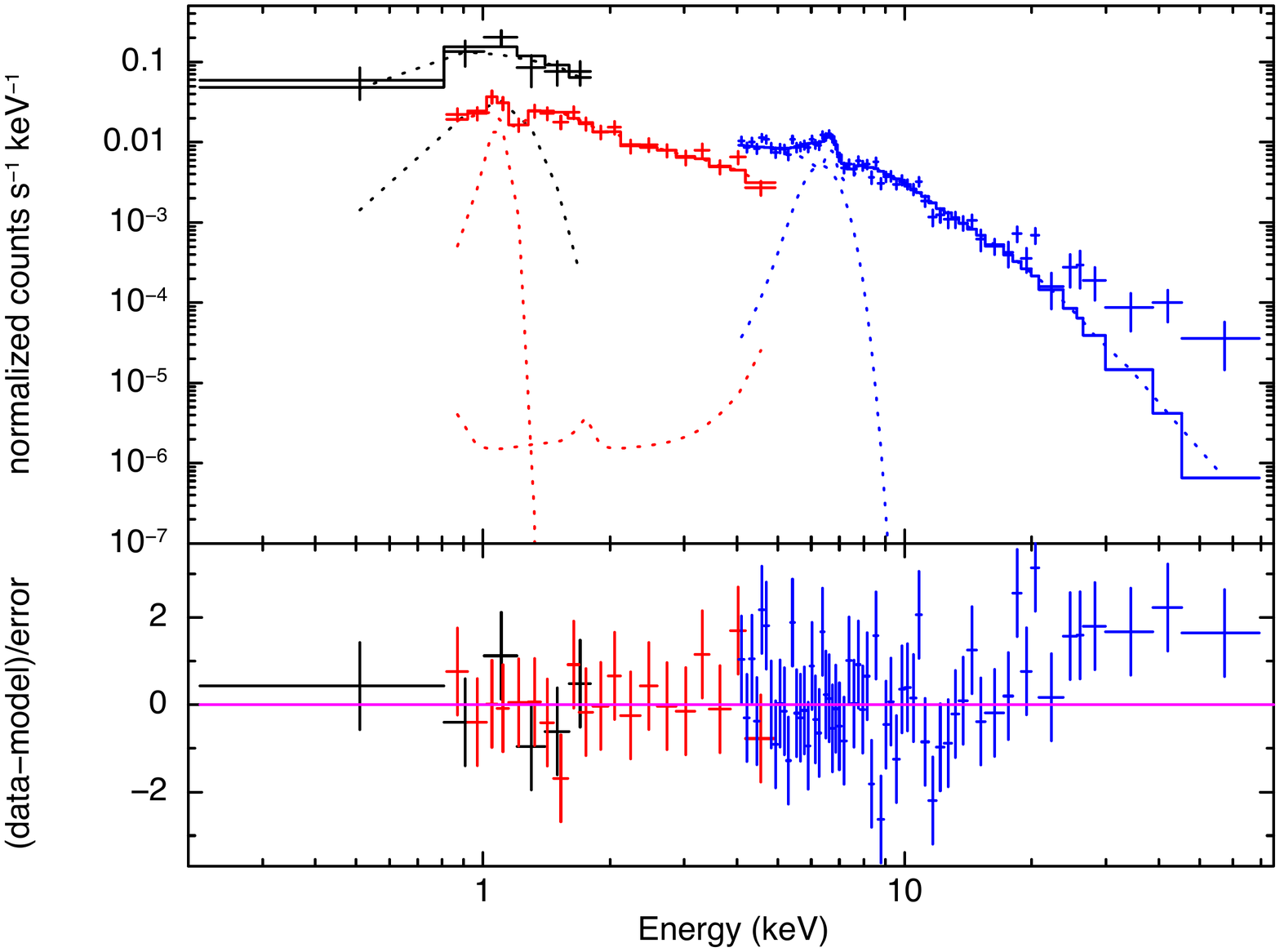} &
\hspace{-0.8cm}\includegraphics[width=2.5in,height=2.5in,angle=0]{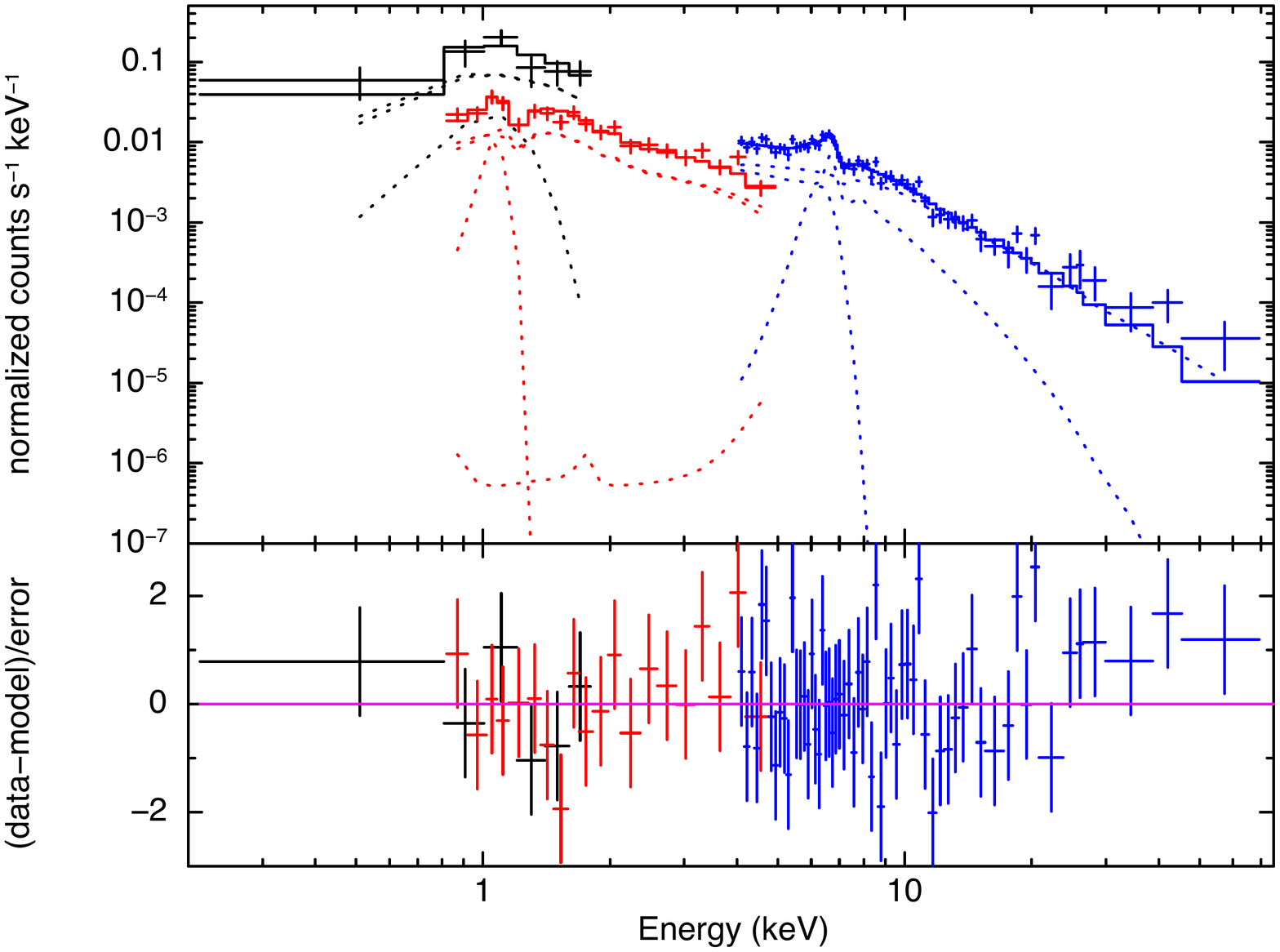} &
\hspace{-0.8cm}\includegraphics[width=2.5in,height=2.5in,angle=0]{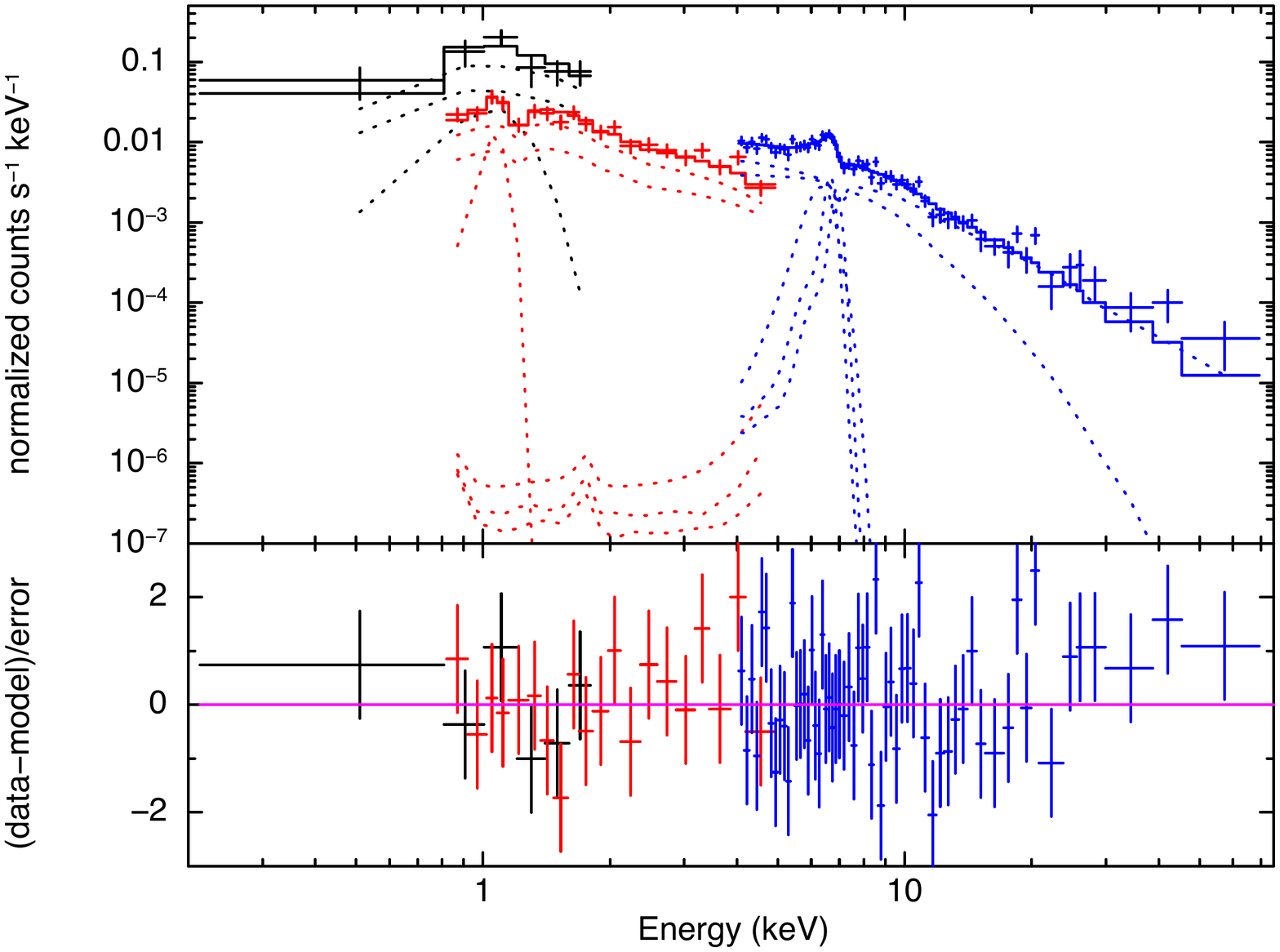}
\end{tabular}
\caption{The left hand panel is the fitted \rosat, \swi, and  \nustar\ (joint) spectra of V592 Cas using the model,
$tbabs$$\times$GABS$\times$GABS(VNEI$\texttt{+}$GAUSS). The middle panel is the joint spectra fitted with the composite model 
$tbabs$$\times$GABS$\times$GABS(VNEI$\texttt{+}$power$\texttt{+}$GAUSS) and the right hand panel shows the same spectra fitted with the model 
$tbabs$$\times$GABS$\times$GABS(BREMS$\texttt{+}$power$\texttt{+}$GAUS$\texttt{+}$GAUS$\texttt{+}$GAUS$\texttt{+}$GAUS) as listed in Table 2. 
The dotted lines show the individual
model components, the crosses are the data (black for \rosat, red for \swi, and blue for \nustar) 
and the solid lines indicate the fitted spectrum. The lower
panels show the residuals in standard deviations.\label{fig3:sp}}
\end{figure*}

\subsection{BZ Cam}

Table 1 shows two composite model fits performed on the joint \rosat, \swi\ and \nustar\ spectra of BZ Cam that describe the conditions in the 
X-ray emitting region. We used Fit-1 that includes a nonequilibrium ionization (NEI) plasma model (VNEI in XSPEC)
since the CEVMKL (CIE model) model fails to fit the composite spectra. Deviations from standard CIE plasma line profiles are expected
in NEI hot flows. Thus, we included emission line and absorption line 
components  to the composite models until the residual fluctuations (2$\sigma$-3$\sigma$) around the lines were reduced. As Fit-1
represents a more physical model for the plasma flow in the X-ray emitting region where only an absorption line of Fe, the iron reflection line at 6.4 keV and 
an absorption feature at around 1.14-1.25 was included for a satisfactory fit.  Fit-2 is utilized to model all the 
detected emission and absorption features with a free-free continuum model that is Bremsstrahlung (BREMSS in XSPEC). We note that
a power law model for the continuum can also be assumed for the same fit but it does not change the modeled line structure
in  Fit-2 and does not significantly improve the fit. The spectral analysis reveals that a  lower plasma temperature in a range 8.2-9.4 keV is obtained with the
VNEI model  instead of 14-19 keV derived from the CEVMKL model fit. The NEI model fits yield lower plasma 
temperatures in comparison with CIE model fits as a result of the plasma ionization conditions (i.e., under-ionized). We note that we have used solar abundances
all throughout the fitting procedures  for both NL sources since the fits are not sensitive to abundance variations where error ranges include solar values.  Overall, Fit-1
improves at 3$\sigma$ significance when a power law model is added to the fit  ($F-Test$ yields improvement at 99.3\% Confidence Level). This
indicates the existence of Compton up/down-scattering in the system. However, Fit-2 is better than Fit-1 at 97.9\% Confidence Level, thus an additional power law model
exists only if NEI plasma is assumed and a fully ionized plasma (as in BREMSS model description) is adequate to explain the continuum emission (however, not the line emission).

In addition, we detect  the iron K$\alpha$ reflection line at around 
6.3-6.5 keV in  Fit-1 and Fit-2. The Gaussian line widths, $\sigma$, are different as calculated by the two fits, where Fit-1 yields 0.2-0.4 keV and Fit-2 0.8$<$ keV, respectively. 
The large width in Fit-2 is indicative of scattering in an accretion disk corona, but the smaller widths in both of Fit-1, portray broadening in the
inner regions of the accretion disc as the flow is extended and has higher viscosity or it is scattering off the existing wind outflow. The Gaussian line sigma for the reflection line 
from Fit-2 and Fit-1 are both inconsistent with Doppler broadening which yield disk radii inside the WD (i.e., too fast a rotation). The mass of the WD (0.4-0.7 \msun)
and the locations where the lines are produced are unlikely for any gravitational redshift interpretation. Thus, the iron K$\alpha$  line widths
are indicative of scattering in the system within the advective hot flow that is extended or the reflection off the cooler parts of the disk, or scattering from the wind outflow itself.  
Reflection off of the accretion disks in AGNs and XRBs are well studied and reveal a Compton reflection spectrum over 0.1-10 keV with a hump around 20-30 keV which is a strong function of the photon index of the incident spectrum, iron abundance, line of sight angle and 
ionization level in the reflected spectrum  \citep[see,][]{2005Ross,2016Kinch}. A reflection broad-band spectrum can be justified for BZ Cam 
owing to the iron reflection line and the additional power law emission (or the large line width with the free-free continuum), however the Compton hump around 20-30 keV is not significant. Fits
using simple  reflection models within XSPEC (e.g., $reflect$) do not produce any better fits than in Table 1 and some reflection models are not suitable for AWBs. 

\subsection{V592 Cas}

Following the spectral analysis of BZ Cam, the analysis of V592 Cas has been conducted in the same manner with the same models (see Table 2) except that this time there is a need for
an additional power law model along with the BREMSS model assumed for the continuum as referenced in the Table 2 comments. We also note that including the H-like iron line in the Fit-2 has low significance  (at only 1$\sigma$), thus there is only a significant He-like iron line which can be derived from the fits.  
The width of the H-like and He-like iron (for both of Fit-2)  
has been kept fixed at around their best fit parameters as they reveal narrow line structure limited with the spectral resolution of \nustar\ at these energies. Fit-1 which is  the VNEI model fit yields plasma temperature in a range 11.1-12.9 keV with an ionization timescale (1.7-3.2)$\times$10$^{11}$ cm\ s$^{-3}$. This timescale is indicative of nonequilibrium conditions. The ionization timescale of BZ Cam is also, (2.9-5.9)$\times$10$^{11}$ cm\ s$^{-3}$ and shows existence of plasma flow that has not reached ionization equilibrium. An additional power law model included in  Fit-1 for V592 Cas is significant at 98.9\% Confidence Level and in Fit-2 is at 99\% Confidence Level which reveals a power law component at high significance. Such an additional component exists significantly for BZ Cam if the VNEI model is assumed. The range of photon indices are 1.5-1.8 for V592 Cas and 1.76-1.87 for BZ Cam. Note that the fitted plasma temperatures become lower once the power law component is included in the fits (see Table 1 and 2). 
V592 Cas shows a 6.2-6.5 keV iron K$\alpha$  line as in BZ Cam. The Gaussian line width, $\sigma$, is 0.2-0.4 keV revealing a similar origin of 
 scattering in the system within the extended advective hot flow or  reflection off the disk, and or reflection off the wind outflow. 
 Similar with BZ Cam, reflection line width shows no consistency with  Doppler broadening or gravitational redshift effects. We note that using simple reflection models as in the analysis of BZ Cam does not improve the
 fits in Table 2.
 
Fit-1 and  Fit-2 show the existence of a Gaussian absorption line (2$\sigma$-3$\sigma$ significance) with a centroid of  6.97-7.13 keV and 6.94-7.05 keV for BZ Cam and V592 Cas, respectively, in the vicinity of  the H-like iron emission line that is at 6.97 keV (FeXXVI). The absorption line widths ($\sigma$) of 0.4-0.08 keV for V592 Cas and 0.1-0.03 keV for BZ Cam, indicate a wider line for V592 Cas and plausibly different conditions in the region of origin and  a different geometry  that is responsible for the absorption.  Furthermore, the fits yield  another absorption feature at around 1.1-1.25 keV in both sources which may correspond to Ne H$\beta$ line or other iron L-shell lines that fall in these energies.   In addition, an emission line is detected in the fits for both of the sources around 0.98-1.03 keV for BZ Cam and 1.00-1.1 keV for V592 Cas which is in accordance with Ne X (H-like Ne) emission line or some iron L-shell emission line. 

Finally, we note that \nustar\ spectra of both of the NLs can be fitted alone with two main spectral components of a thermal plasma and a power law yielding power law photon indices in a range 1.85-2.2. The significant power law components from the \swi\ analysis are already mentioned in Sec.~\ref{int:xray}.  The neutral hydrogen column density in the line of sight for BZ Cam and V592 Cas was throughly discussed in \citet{2014Balman} and will not be repeated here, but the values derived from the fits given on Tables~\ref{tab:sp1} and ~\ref{tab:sp2} are consistent with absorption by the 
interstellar medium and we do not find any intrinsic neutral hydrogen column density that could yield absorption of any alleged soft X-ray emission component.
 
\section{Aperiodic variability in the accretion flow}\label{sec:pds}

Analysis of aperiodic  variability and broadband noise have been used as a diagnosis technique for understanding disk structure and state transitions in accreting binaries. 
Long timescale variability in CVs disks may be created in the outer parts of the accretion disk, but the relatively short-time variability
(at $f>$few mHz), intrinsic to the disk originates in the inner parts of
the accretion disk \citep[see][and references therein]{2012Balman,2012Scaringi,2012Dobrotka,2015Balman,2015Bruch,2016Baptista,2016Dobrotka,2019Balman}.
Properties of this noise are similar to that of the X-ray binaries with neutron stars and black holes.
The model of origin for this aperiodic noise is the model of propagating fluctuations \citep{1997Lyubarskii,2009Revnivtsev,2010Revnivtsev,2011Uttley,2011Ingram,2013Ingram,2016Ingram}.
Variations in the mass accretion rate as a result of fluctuations, are inserted into the flow at all Keplerian radii of
the accretion disk due to the nature of its viscosity and then transferred toward the compact object. 
In this prescription, optically thick accretion flows are expected to show  a frequency index of $f$$^{-1\ldots -1.3}$ in the
power spectral distribution of the red noise \citep{2001Churazov,2005Gilfanov,2010Gilfanov}. 
Once the optically thick flow subsides (truncates in a region/radius), the broadband noise as a result of aperiodic variability of the flow which has a power law ($\sim$1/f) distribution in frequencies will show a break (red noise which shows the 
characteristic of the optically thick flow variability will show a break from the 1/f dependence). It is expected that red noise will diminish as a result of 
lack of aperiodic variability from the optically thick flow. This red noise (power law) will show a break into a steeper (e.g., $1/f^2$ \citealt{2006Klis}) power law regime in higher frequencies. Unless otherwise another noise component takes over.
Historically, truncation of the optically thick accretion disk in DNe in
quiescence was introduced as an explanation for the time lags between the optical and UV
fluxes in the rise phase  of the outbursts \citep{1994Meyer,1999Stehle}, and for some implications of the Disk Instability Model (DIM) \citep{2004Lasota}. 

CVs demonstrate band limited noise (noise structure that steeply decreases or shows a break towards high frequencies) in the optical, UV, and X-ray energy bands,
which can be adequately explained in the framework of the 
model of propagating fluctuations. Detailed modeling and results can be found in \citet{2012Scaringi,2012Balman,2015Balman,2016Semena,2019Balman}. 
The detected frequency breaks in the nonmagnetic CVs (mainly DNe) are 
in the range (1-6) mHz in quiescence and indicate an optically thick disk truncation (i.e., transition) showing existence of 
the advective hot flows (RIAF ADAF-like flows) in the inner regions. 
Analysis of available data (e.g., SS Cyg, SU UMa, WZ Sge) reveal that during the outburst the inner (optically thick) disk radius moves towards 
the white dwarf and recedes as the outburst declines \citep{2019Balman} while linked changes in the X-ray energy spectra are also observed. 
Cross-correlations between the simultaneous optical, UV and X-ray light curves show time lags (propagation lags about 1.5-3 min) consistent with truncated optically thick disk and transition to advective hot flows in the inner disk \citep[see][for a final review]{2020Balman}. The aperiodic variability characteristics of NLs have not been studied in detail (using the propagating fluctuations model) 
except for MV Lyr \citep{2012Scaringi,2013Scaringi} in the optical revealing several QPO structures and a break in the frequencies roughly around 1 mHz.  
Here, we study the power spectral characteristics using the \nustar\ data of 
both BZ Cam and V592 Cas and include the analysis of MV Lyr for comparisons.

In order to study the noise structure of the power spectra (PDS), we prepared source and background light curves with the $nuproducts$ task  using the standard procedures of \nustar\ analysis for each source, MV Lyr, BZ Cam, and V592 Cas as discussed in Sec. 2. The background subtracted light curves were calculated using  FTOOLs and  power spectral analysis were done utilizing the {\sc XRONOS}\footnote{https://heasarc.gsfc.nasa.gov/lheasoft/ftools/xronos.html}\ software package and the {\it powspec} task as part of HEASoft. During the timing analysis, the background subtracted FPMA/B light curves were used simultaneously.
PDS were calculated in terms of the fractional rms amplitude squared per Hertz following the \citet{1991Miyamoto} normalization \citep[see also][]{1990Belloni}. While calculating final PDS, light curves were divided into segments using 1-2 sec binning in time and a total bin number per PDS, several PDSs were averaged (taking into account observation windows).  The white noise level was subtracted from each averaged PDS leaving the rms  
fractional variability of the time series and the powers were multiplied with the frequencies yielding $\nu P_\nu$ prescription (i.e, integrated power in (rms/mean)$^2$; \citet{1997Belloni}). 
Our resulting PDSs for our sources are displayed in Figure 4.  
We do not find any effect of the orbital periods for any of our three sources (0.075-0.1 mHz)  in the PDSs we calculate. This is an effect of our analysis method (stack-averaged PDS) and the fact that these periodicities are in the error range of the lowest frequency bin of our calculated PDSs. In addition, we do not have a contribution (i.e., periodicity) from the satellite (\nustar) orbital motion ($\sim$ 0.17 mHz) for the same reasons. These periodicities are long for the time segments we use in the analysis.
Subsequently, we fitted the PDS with the analytical model $\nu P(\nu) =  P_2\times(1+(\nu/P_1)^4)^{-1/4}$  (where P$_1$ is the break frequency $\nu_0$) that is used to describe the PDS of sources with truncated accretion disks \citep{2010Revnivtsev,2011revnivtsev,2012Balman,2015Balman}.  Additional Lorentzian functions were introduced for adequate modeling of the PDS.
The fits show frequency breaks in the PDS of our NLs as  3.9$\pm$2.5 mHz for MV Lyr, 2.5$\pm$2.0 mHz for BZ Cam, and $\le$10 mHz for V592 Cas (ranges for BZ Cam and MV Lyr  correspond to 90$\%$ Confidence Level and the \rchisq of the fits are 0.9-1.3).  During the fitting procedure, one Lorentzian with a  peak frequency at about 1.7$\pm$0.7 mHz was included for BZ Cam which compensated  for the peaked noise component however, the significance of this component is  only 1$\sigma$. For V592 Cas, two Lorentzians were used
to achieve a good fit. One of them is a QPO component (i.e., narrow, [$\nu$/FWHM]$<$2 \citealt{2006Klis}) with a peak at 1.4$^{+2.6}_{-0.3}$ mHz and the other is a peaked noise component ([$\nu$/FWHM]$>$2) with 3.6$\pm$0.4 mHz (ranges of errors correspond to 90$\%$ Confidence Level).  The narrow QPO component is detected in both the power spectrum
with the Miyamoto normalization and its integrated power spectrum ($\nu P_\nu$). Our assumed broken power law model was fitted together with the Lorentzians to find the break frequency.  However, the fits were not sensitive to this component and we derived an upper limit of  $<$10 mHz for V592 Cas.  Our results show that the PDS structure of these three NLs resemble DNe PDSs and its variations between quiescence and high states which will be elaborated in the next section.
 
\begin{figure*}
\begin{tabular}{lll}
\hspace{-1.0cm}
\includegraphics[width=2.5in,height=2.5in,angle=0]{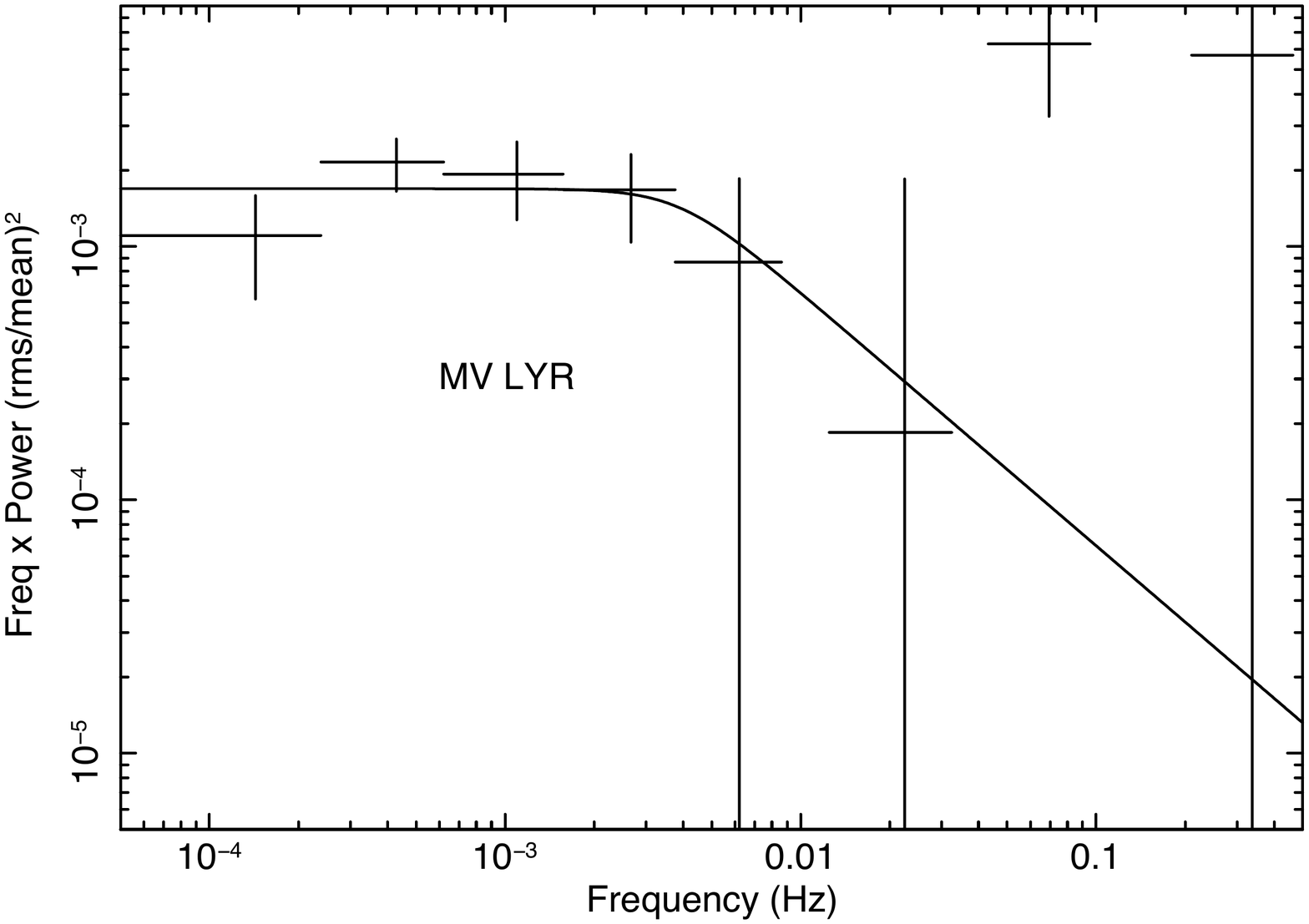} &
\hspace{-0.8cm}\includegraphics[width=2.5in,height=2.5in,angle=0]{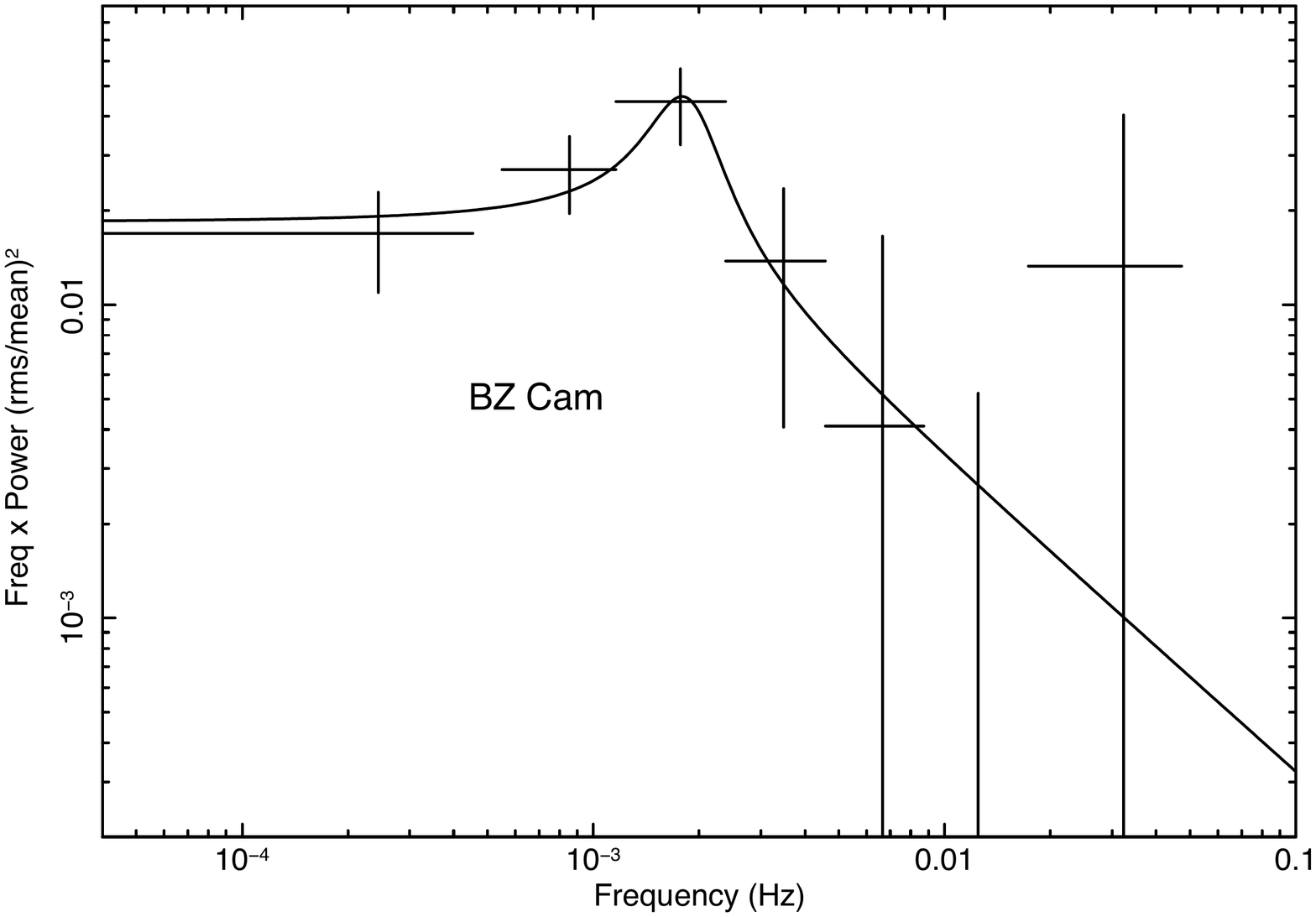} &
\hspace{-0.8cm}\includegraphics[width=2.5in,height=2.5in,angle=0]{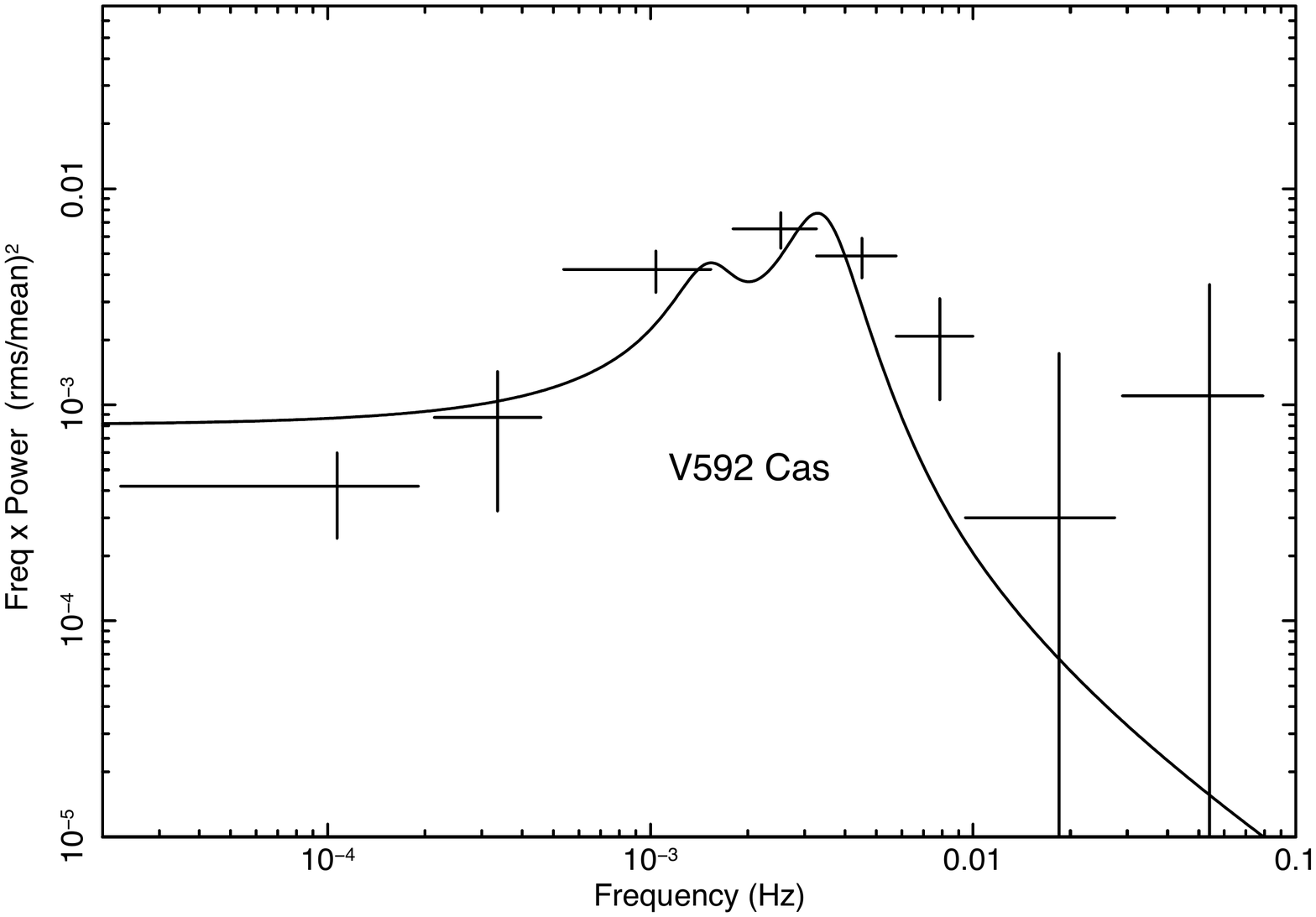}
\end{tabular}
\caption{The power spectra (PDS) of MV  Lyr (for comparison), BZ Cam, and V592 Cas obtained using \nustar\ light curves. Overplotted are the fits with the
propagating fluctuations model for  MV Lyr, BZ Cam, and V592 Cas. One Lorentzian for BZ Cam and two additional Lorentzian models for V592 Cas were used  to achieve an adequate fit.\label{fig4:sp}}
\end{figure*}

\section{Discussion}\label{sec:discuss}

\subsection{The power spectra (PDSs)}

We have investigated the red noise structure in three NL sources (BZ Cam, V592 Cas, and MV Lyr for comparison) to reflect upon the advective hot flows independently from the spectral analysis, study their variability characteristics,  and find their transition radii where the optically thick flow truncates in NL type CVs.  Variations in the mass accretion rate are inserted into the flow at all Keplerian radii of an accretion disk and then transferred toward the WD, thus variations are on dynamical timescales. Propagating fluctuations model predicts that a truncated optically thick accretion disk should lack some part of its aperiodic variability at high Fourier frequencies. The calculated
PDSs show frequency breaks at  3.9$\pm$2.5 mHz for MV Lyr, 2.5$\pm$2.0 mHz for BZ Cam, and $<$10 mHz for V592 Cas with a peaked noise component at 1.7$\pm$0.7 mHz for BZ Cam and 3.6$\pm$0.4 mHz for V592 Cas. Assuming the simple relation, $\nu_0 = 1/2\pi (GM_{WD}/R_{in}^3)^2$, the frequency breaks correspond approximately to a transition/truncation radii at about  R$_{tr}$ $\sim$ 5.5$\pm$2.0$\times$10$^9$ cm for MV Lyr, $\sim$ 7.4$\pm$2.4$\times$10$^9$ cm for BZ Cam, and $>$3.0$\times$10$^9$ cm for V592 Cas. This is a strong indication of extended structures fort our NLs in the X-ray regime. 
In addition, we have detected a QPO at 1.4$^{+2.6}_{-0.3}$ mHz  for V592 Cas where an optical QPO had been detected previously, as well, at a lower frequency of about 0.75 mHz \citep{2002Kato}. 
Figure 4 displays the PDS of MV Lyr, BZ Cam and V592 Cas with an increasing accretion rate from left to right. These three sources (NLs) are high state CVs, but MV Lyr indicates a PDS more similar to quiescent DN at the low accretion limit of  about 10$^{-9}$\msun\ yr$^{-1}$ calculated from UV/optical. As seen from the figure increasing accretion rate shifts the noise towards higher frequencies and in BZ Cam we see a peaked noise component most likely originating in the  advective hot flow (increased variabilty in the hot flow).  This peaked noise component originating from the hot flow perhaps becomes more coherent once the accretion rate surpasses 10$^{-8}$\msun\ yr$^{-1}$ yielding the QPO structure we detect and the peaked noise moves towards a higher frequency.  It has been suggested that such a QPO may be associated with the break frequency or 
have to do with the precession/oscillations of the inner hot flow \citep{2006Klis,2013Ingram}.  We note that V592 Cas has a precessing disk with positive and negative superhumps (see Sec.~\ref{sec:NLs}). 
The PDS of V592 Cas resembles the PDS of SS Cyg             
at high state (during the optical peak) with a similar accretion rate \citep{2007Schreiber}, however the truncation radius is further out in the disk of V592 Cas about by at least factor of 3.  As for the correspondence of PDS with other X-ray binaries, 
a long discussion has been made in \citet{2019Balman} for general CVs and the changes of the noise structure with increasing accretion rate is observed in other X-ray binaries with noise shifting to higher frequencies and becoming more coherent (appearence of QPOs), and Lorentzian peaks are used to model the hard state or quiescent PDS of LMXBs where the advective hot flows dominate (the variability of) the accretion disk \citep{2010Belloni,2014Munoz-Darias,2018Degenaar} .

\subsection{Characteristics of Advective hot flows}

The X-ray emission from NLs and nonmagnetic CVs (and perhaps most AWBs) are adequately explained with radiatively inefficient advective hot accretion flows (RIAF ADAF-like) in the close vicinity of the 
WD \citep[see][and references therein]{2015Balman,2020Balman}. We find that the RIAF ADAF-like accretion flows in nonmagnetic CVs are mainly thermal in X-rays with low or higher nonequilibrium 
ionization effects due to the nature of the flow, its viscosity and its turbulence. One of the basic property of such flows is the radiative inefficiency which we calculate in this study within a wide energy band 
of 0.1-100 keV for both BZ Cam and V592 Cas. The spectral results in Tables 1 and 2 show a total X-ray luminosity of  (1.1-1.6)$\times$10$^{32}$ \lumcgs\  and  (1.2-1.8)$\times$10$^{32}$ \lumcgs\ for BZ Cam 
and V592 Cas, respectively. The accretion luminosities of these sources as calculated from the optical and UV wavelengths are $\sim$ 3$\times$10$^{34}$ \lumcgs\ and $\sim$ 1.2$\times$10$^{35}$ \lumcgs\ 
(V592 Cas) \citep[cf. Table 1][and references therein]{2014Balman}. As a result, the radiative efficiency ($\epsilon = L_x/L_{opt/UV}$) in the X-ray emitting region is $\sim$ 0.004-0.005 for BZ Cam 
(an upper limit since L$_{opt/UV}$ can be higher) and $\sim$ 0.001-0.002 for V592 Cas. These calculations of luminosities (i.e., larger than \swi\ results) and
efficiencies (i.e., smaller than \swi\ results) are consistent, but are revised versions of the \swi\ spectral results presented in \citet{2014Balman}. Therefore, X-ray emission efficiency is about 0.1\%-0.5\% of
the accretion luminosity of these sources.  

The nonthermal power law emission components detected in this study, have luminosities 
of (4.6-5.4)$\times$10$^{31}$ \lumcgs\  (BZ Cam)  and (0.9-1.1)$\times$10$^{32}$ \lumcgs\  (V592 Cas) and are a factor of 2 larger for V592 Cas.
Nonthermal (power law-type) X-ray emission is about 0.2\% of the accretion luminosity for BZ Cam and it is about 0.1\% of the accretion luminosity of V592 Cas. In general,  inefficiencies in the X-ray luminosities are consistent with expectations of radiatively inefficient advective hot flows (RIAF ADAF-like) in the X-ray emitting regions. The existence of low luminosity power law emission in the X-rays is justified since the accretion flow does not cool  either thermally or via Compton cooling (i.e., marginal cooling). As the flow does not radiate and cool, the energy is carried within the flow and advected into the WD heating the WD or  energy is channeled to aid formation of outflows from the disk in the X-ray emitting or other regions further out \citep[see also][]{1995Narayan}. 
 
\citet{2008Narayan} and \citet{2014Yuan}  review the characteristics of advective hot flows in several different regimes, mostly for black holes.  One of the basic criteria is the efficiency factor which is given with respect to the 
Eddington luminosity prescription as $\epsilon$=$L/(\dot{M}_{BH} c^2$) (where L is the source luminosity and the denominator represents total accretion energy budget of the black hole). 
In the regime $\dot{M}/\dot{M}_{Edd}$ $<$10$^{-3}$ (e.g., depends on the square of the $\alpha$ parameter) advection dominates the flow and as the accretion rate increases, the fraction 
of advection in the  flow drops and the flow becomes more radiative. 
Note that fraction of advection is a function of $r$ in the disk and the accretion rate, and  the optically thick part of the disk flow can be sustained. This yields a two-zone accretion flow where there is a cool geometrically thin disk at 
large radii and a hot accretion flow at small radii \citep[][and references therein]{2007Done}.  In addition, there is a TCAF model (two-component ADAF model) where the optically thick Keplerian disk is surrounded by a sub-Keplerian flow (halo) and a standing shock leads to an inner extended post-shock halo \citep{1995Chakrabarti,1990Chakrabarti}. 

At low $\epsilon$ (efficiency)  limit gas is heated but hardly cools (our NL sources resemble this). In this regime both electrons and ions are radiatively inefficient, electrons 
are unable to radiate the viscous heat they acquire and the Coulomb collisions are inefficient for heat transfer.  Note here that radiative cooling depends on gas density, thus this is also low.
As $\dot{M}$ increases,  $\epsilon$ also increases where electrons start radiating  the viscous energy they attain and part of the energy they obtain from ions \citep{2014Yuan}. 
These type of advective hot flows (RIAF ADAF-like) at relatively lower accretion rates compared to $\dot{M}_{Edd}$ are expected to be in nonequilibrium ionization creating 2-T flow structure.  
As a result, the existence of NEI flow structure is a typical characteristic and we recover this for both of the NL sources in our study.  Furthermore,  
Comptonization of the flow depends strongly on accretion rate and is expected to be a weak component at low accretion rates  and at low radiative efficiencies compared to Eddington values.  The power law components 
of BZ Cam and V592 Cas (2 times stronger than BZ Cam in luminosity) are only (0.1-0.2)\% of their accretion luminosity  which are consistent with these expectations.
Moreover, in ADAF flows at radii where the gas is cool enough, heavier atomic species, iron peak elements can retain electrons and show emission lines \citep{1999Narayan}.  
Note here that RIAF  ADAF flows are virialized with expected temperatures of 
MeVs in BH or NS compact binaries, but the X-ray emission is detected at temperatures around 100 keV or less owing to a 2-T regime between electrons and ions and/or  several cooling mechanisms that exists in different hot flow regimes \citep{2014Yuan}.  However, the virial temperatures in AWBs (i.e., for CVs) are 10-45 keV \citep{2014Balman,2020Balman} and the observed X-ray temperatures are 
consistent with virial temperatures, therefore the flows are already virialized.  On the other hand, the X-ray temperatures of the two NLs in this study are somewhat lower due to the NEI  characteristics of the flow 
(fitted models reflect the electron temperatures).  We stress here that ADAF flows that represent a 2-T flow may be approximated with a single T (flow may mimic single T) 
as these are not the type of 2-T flows around BHs since they are in a shallower gravitational potential well with no GR effects. 
However, effects of turbulence due to magnetic effects (i.e. Alfvenic waves) and MRI related turbulence (magnetorotational instability) should also be considered for advective hot flows. In general, magnetic turbulence promotes disequilibration between electrons and ions. When magnetic energy density is greater than thermal energy density in the flow, electrons are preferentially heated whereas in the opposite case, ions are heated \citep[][and references therein]{2010Howes,2019Kawazura}. Even at constant magnetic field in the flow, turbulent heating would gradually increase the thermal to magnetic energy ratio of ions to a state of dominant ion heating. Thus, existing magnetic fields and turbulence are capable of pushing a weakly collisional plasma system away from thermal equilibrium. 
 
Recently,  \citet{2021Datta} have performed calculations of shock formation in ADAF flows around accreting WDs. For a transonic flow around an accretor with a hard surface (i.e., WD or NS) shocks are to occur given some boundary conditions and system parameters as the flow has to slow down at the surface \citep{2016Dhang}. Formation of shocks enhances in a transonic sub-Keplerian flow once inner and outer critical points co-exist \citep{1996Chakrabarti}. \citet{2021Datta} solve the hydrodynamical problem between these critical points to see when the shock conditions are satisfied and find that for a WD of mass 0.8\msun\ and 0.01\rsun\ , and for other given system parameters like rotation and eccentricity, a shock occurs around 1.3$\times$10$^9$ cm which is very close to the WD.  Shocks, in general will enhance densities, hence coulomb collisions and emissivity and also enhance radiative efficiency.  Formation of shocks in the advective hot flows can be articulated in the context of all AWBs. Given section 2.3 of \citet{2020Balman}, the advective hot flows in quiescent dwarf nova mimic CIE flows with near solar abundances and inefficiency in line emissivity. DNe in quiescence
show optically thick disk flow truncation of radii (3-10)$\times$10$^9$ cm \citep{2012Balman} which yields enough distance for acceleration of the transonic advective hot flow and formation of shocks near the WD \citep[see also][]{2021Datta}. This will yield enhancement of X-ray emission (increase of efficiency) in a small zone near the WD which can also explain some of the detected eclipse-like effects at high inclinations (rather non-detections at low count rates) in quiescent DNe. Existence of shocks can be effective in NLs, as well.

\begin{figure}
\centerline{
\includegraphics[width=2.7in,height=2.7in,angle=0]{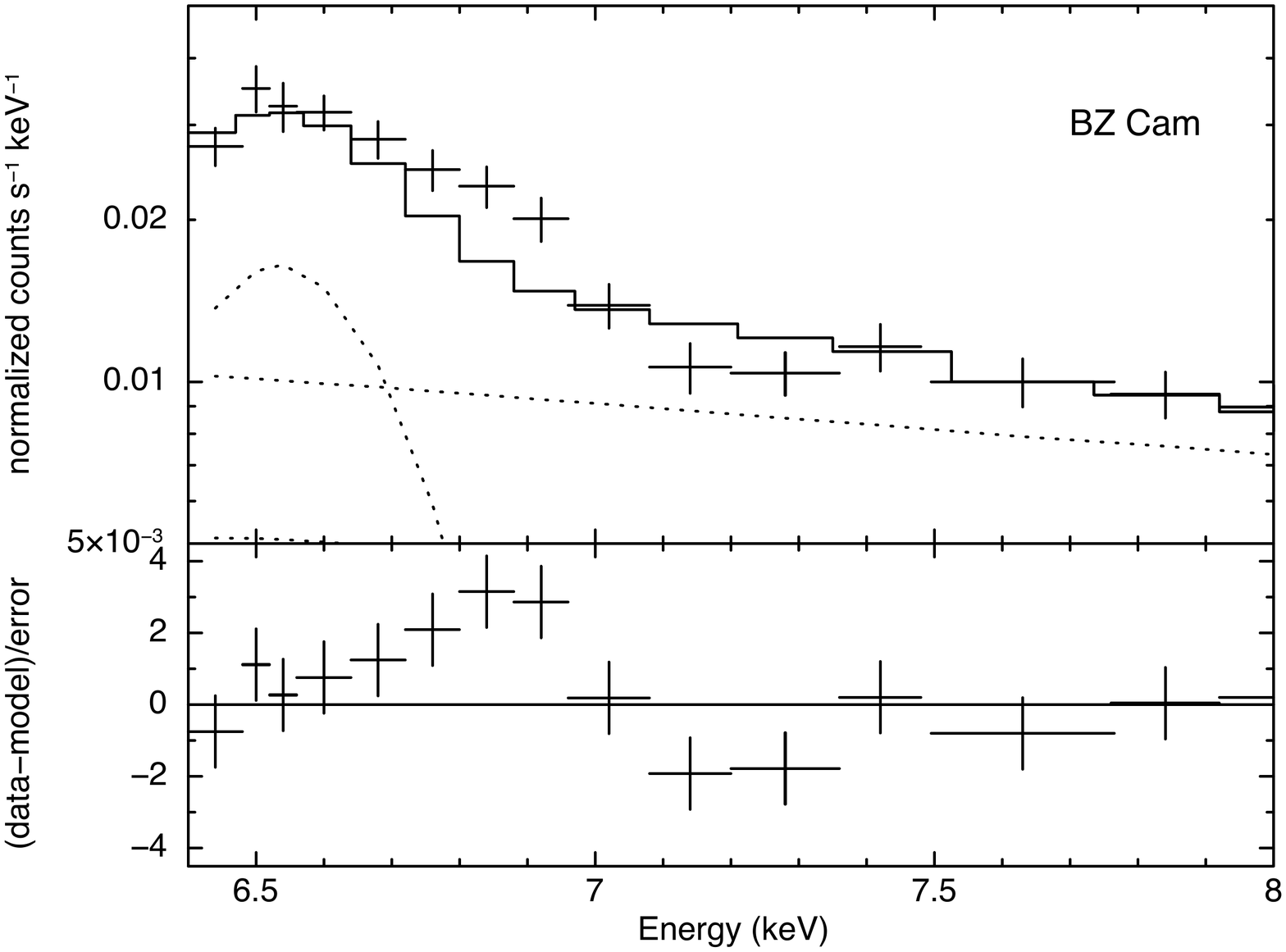} 
\includegraphics[width=2.7in,height=2.7in,angle=0]{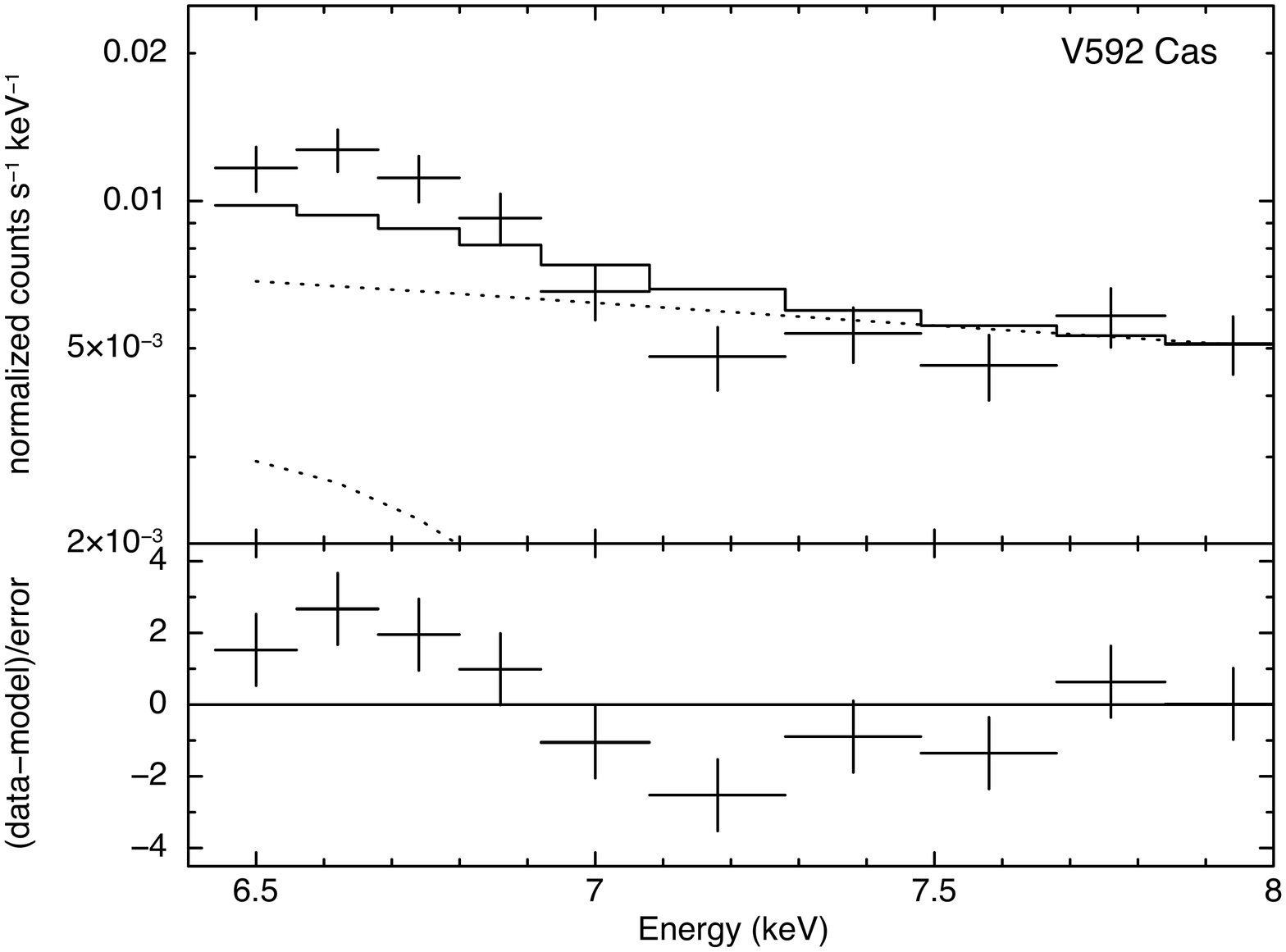}}
\caption{The variation around the H-like iron line for BZ Cam on the left and V592 Cas on the right. The fitted spectrum of BZ Cam
using model Fit-2 in Table 1 is displayed where only the H-like iron emission line and the absorption feature in this energy are excluded from the fitting process. Note the
P Cygni-type variation in the residuals (lower panel) around the expected ionization line centroid 6.97 keV.
On the right are the deviations in sigmas (residuals) produced using Fit-2 for V592 Cas in the same manner with BZ Cam along with an additional exclusion of the He-like iron line for V592 Cas. The figure mostly indicates the significance of the He-like line and the H-like absorption line. No P Cygni-type variation as in BZ Cam is observed. \label{fig5:sp}}
\end{figure}

\subsection{Iron Emission and Absorption lines}

The emission lines incorporated in Fit-2 of both sources are designated iron ionization emission lines of He-like (Fe XXV) and H-like (Fe XXVI) species expected from the thermal plasma along with the assumed Bremsstrahlung continuum. These lines are detected for BZ Cam at around 6.88-6.91 keV (H-like) and 6.52-6.57 keV (He-like) indicating a NEI nature for the accretion flow in the X-ray emitting region (ranges correspond to 90\% Confidence Level). V592 Cas shows dominantly the He-like emission line (complex) at around 6.56-6.66 keV. The H-like ionization line is included for completeness and detected at 1$\sigma$ significance around 6.84-6.96 keV where the energy ranges and dominance of the He-line (absence of the H-like line) support the NEI nature, as well.  As the plasma in these NLs is not in equilibrium,  the line centroids from the under-ionized plasma do not appear at their equilibrium values (6.97 keV for the H-like and the 6.7 keV for the He-like iron emission lines), and they are detected at slightly lower energies. As a result,  Gaussian emission lines used for Fit-2 on Table 1 and 2 reveal NEI conditions on the lines and the plasma, so Fit-2 and Fit-1 (NEI model fits) confirm each other explicitly. 
Note that the He-like line is a blended triplet line feature that is composed of the resonance, intercombination and the forbidden line emission components (due to inadequate spectral resolution). The NEI conditions
imply shifts to lower energies from the equilibrium line centers is a result of the fact that the forbidden line emission is enhanced  in the triplet emission.  Given the low hydrogen column densities derived
from  the fits for both NLs, if the iron line emission was due to photoionization, the resonance line should have dominated the He-line complex and this is not what we observe \citep[see][]{2005Bianchi}.  The range of ionization timescales  are  
(2.0-9.0)$\times$10$^{11}$\  s\ cm$^{-3}$ (within 90\% Confidence Level) where these values indicate that the plasma is not in equilibrium as the time scales would then exceed several times  10$^{12}$\  s\ cm$^{-3}$ \citep{1999Liedahl}.  We note also that a study published on the archival \cha\  HETGS data of CVs by \citet{2014Schlegel}  indicate line emission consistent with multi-temperature plasma in a nonequilibrium state with n$_{e}$ between 10$^{12}$-10$^{16}$ cm$^{-3}$ that can be attributed to characteristics of advective hot flows. 

In order to reduce the residual fluctuations around the iron line complex further modeling showed 
an absorption line around 7.00-7.13 keV for BZ Cam and  6.94-7.05 keV  for V592 Cas with significant widths (see \ref{sec:spec}). These absorption lines indicate
optical depths at the line centers of 1.13-0.7 for V592 Cas (i.e. optically thin flow), but for BZ Cam the ranges change from 4.6-11.6 to 35-62 depending on the model fit. We believe the inconsistency for BZ Cam is a result of the P Cygni structure on the line shape which we will elaborate in the following paragraph. 
Such iron absorption lines are indicators of warm absorbers and thus, their detection suggests existence of warm absorption in both of the NLs similar to X-ray binaries \citep{2009Balman,2016Diaz-Trigo}. 
Both of our sources have substantial wind outflows and existence of extended structure in UV/optical wavelengths (see \ref{sec:intro} and sec. 4 in \citet{2020Balman}).  Note that the iron absorption line ranges are close to or include the iron K-edge absorption around 7.11 keV, however the X-ray spectra and temperatures detected in these two NLs does not adequately support such a scenario for a fully ionized hot accretion disk coronae. 

These iron absorption  line features along with the emission lines at the same energies as the H-like species can be a result of P Cygni profiles in the X-rays 
for the H-like iron emission line. This shows existence of outflows from the X-ray emitting plasma. The outflow speeds can be inferred from the difference 
between the emission peak and the absorption dip energies using the fits in Table 1 and 2.  
This has been readily tested for BZ Cam since the H-like iron line emission  is securely detected (3$\sigma$-4$\sigma$). Both of Fit-2 (also tested with Fit-1) are performed without the emission and absorption lines around the H-like iron line energies. Figure 5 shows the P Cygni-type variation in the residuals for BZ Cam when these two features are removed from Fit-2 with a power law component (everything else remains; the He-like iron and the 6.4 keV reflection line is fitted). The difference of the emission and absorption line centroids (GAUSS and GABS models) in Fit-2 of BZ Cam yield speeds 
(using Doppler formula) 0.029$c$-0.015$c$ that corresponds to 4500-8700 km s$^{-1}$ including the error range of the centroid parameters. This is the first time we are recovering a jet-like outflow in the X-rays from an accreting CV.
The wind speed in BZ Cam range in 3000-5300 km s$^{-1}$, thus the broadband X-ray spectrum reveals  similar or larger outflow speeds as they are produced deeper in the potential well where the inner 
advective hot flow dominates. Note that Figure 5 includes a fit to V592 Cas spectra (using Fit-2 in Table 2) produced in the same manner, excluding the He-like iron line, as well,  for comparison. The existence of a P Cygni-type variation can not be properly confirmed on any H-like iron line.  The figure mostly indicates the significance of the Fe XXV (He-like iron) and the H-like iron absorption line ($\sim$ 3$\sigma$).  
Such characteristic outflows as in BZ Cam (for that matter in V592 Cas assuming optical and UV) have been observed in AGNs and XRBs as disk-driven winds \citep{2000Elvis}, thermally driven winds emanating from the dusty torus of 
AGNs \citep{2001Krolik} and as part of the hot accretion flows and their outflows modeled with MHD/GRMHD \citep[see][for a review]{2014Yuan}. 
The blue-shifted iron K-shell lines in radio quiet AGN have been interpreted with ultra-fast 
outflows (UFO) defined as highly ionized winds (ionization parameter log($\xi$) 3-6) outflowing at velocities in excess of 10000 
km s$^{-1}$ \citep{2010Tombesi}. Thus, UFOs are disk-winds close to the black holes, whereas the warm absorbing disk-winds in AGNs and XRBs are 
slower 300-4000 km s$^{-1}$ and produced further out forming a stratified wind \citep[][and references therein]{2012Ponti,2016Diaz-Trigo}. These winds can carry significant amount of mass comparable or even larger than the amount of accreted material \citep[e.g.,][]{2014Degenaar}.  

The outflow of BZ Cam detected in the X-ray regime can also help to explain the  existing bow-shock nebula and the mass associated with it (see~\ref{sec:intro}).  The bow-shock nebula around BZ Cam is claimed to be of several recurrent novae that occurred in the system with timescales of milenia \citep{2020Hoffmann} that indicates 3-4 shells assuming nova expansion velocities (the detected 
curvature radii of the shells around the source are 16$^{\prime\prime}$-61$^{\prime\prime}$). There are prominent O{\sc III} and H$\alpha$ emission structures  associated with this bow-shock nebula (or the nova shells) \citep{2018Bond}.  \citet{2020Tappert} find H$\alpha$ luminosity of 5.1$\times$10$^{30}$ \lumcgs\  and O{\sc III} luminosity of 9.1 $\times$10$^{30}$ \lumcgs\  where these luminosities overlap within errors. Such a bow-shock like nebula exists around the black hole binary Cyg X-1 (at relatively larger size, but concentric ring-like emission regions  trailing the shock region  (as in BZ Cam) are not detected). It is calculated that this bow-shock nebula is powered by the relativistic jet interaction with the circumstellar medium \citep[see][and references therein]{2007Russell}.  There are prominent O{\sc III} and H$\alpha$ emission structures  associated with this nebula with luminosities of  (1.3-21)$\times$10$^{34}$ \lumcgs\  and (1.8-25)$\times$10$^{34}$ \lumcgs\ in
O{\sc III} and H$\alpha$, respectively. These luminosities  match within errors and are calculated to be a result of a jet power of (4-14)$\times$10$^{36}$ \lumcgs from the binary with X-ray luminosity of
about 1$\times$10$^{37}$ \lumcgs. The nebular luminosities (O{\sc III} and H$\alpha$) of around BZ Cam are about 10$^5$ less than Cyg X-1 and the source X-ray luminosities are less by about the same factor 10$^5$.
Thus, a jet-like collimated fast outflow that could power the BZ Cam bow-shock nebula (with a shock that will mimic a nova shell interaction) should have several $\times$10$^{31}$ \lumcgs in the X-rays scaling from the jet-power in Cyg X-1 nebular interaction. Such X-ray luminosities are well within the detected values in our study and available in the accretion flow since it is radiatively 
inefficient and most of the accretion energy ($\sim$ 1$\times$10$^{34}$ \lumcgs) is carried in the flow (very low efficiency of emission in the X-ray region).  Therefore, we suggest that the
jet-like outflow detected in this study may create and/or power the bow-shock nebula of BZ Cam. The several ring-like structures may possibly be a result of turning-on and off or enhancement in  matter ejection of the jet-like outflow 
within several hundred years timescale. Assuming  mass loss rate in the jet-like outflow is 0.01$\%$ of the accretion rate ($\sim$5$\times$10$^{-9}$ \msun), for a timescale of 10$^6$-10$^7$ yrs  ($\sim$100 times less than CV evolution timescales), 
this will  gather (5-50)$\times$10$^{-5}$ \msun\ mass in the circumstellar medium of BZ Cam which is consistent with standard classical nova ejecta masses \citep{2020Balman} that could mimic a nova remnant and its interaction.

The P Cygni-type profile of the H-like iron line of BZ Cam indicates that we are looking into the outflow, where perhaps the nondetection of such a structure in V592 Cas may mean that the outflow is there, but not directed into our line of sight. The error range of the centroid of the detected H-like absorption line for V592 Cas is consistent with blueshifts (in absorption) that is a result of about 3000 km s$^{-1}$ outflows (using Fit-2) in accordance with the bipolar winds in this system. Thus, existence of a  jet-like outflow from V592 Cas is not justified.
  
 \section{Summary}
 
 In this paper, we have presented the analysis  of the broadband spectra of two NL systems, BZ Cam and V592 Cas, using the \rosat, \swi\ and \nustar\ observatory archival data together with the power spectral analysis of the \nustar\ data. 
 We have considered the nature of the accretion flow and outflows in these systems as observed in the X-ray regime. We show that these objects have a hard X-ray spectrum over the 0.1-78 keV energy range with no soft (black body model) X-ray emission. We find that the X-ray energy spectral and power spectral characteristics reveal radiatively inefficient advective hot accretion flows (RIAF ADAF-like) in these two systems.  Fits with multitemperature CIE model yield
nonphysical flat power law indices showing marginal cooling and emission in the X-ray emitting region. We find X-ray luminosities of (1.1-1.6)$\times$10$^{32}$ \lumcgs\  and  (1.2-1.8)$\times$10$^{32}$ \lumcgs\ for BZ Cam 
and V592 Cas, respectively. The radiative efficiencies for the NLs  are about
 0.004-0.005 for BZ Cam and 0.002-0.001 for V592 Cas comparing their X-ray luminosity with the accretion luminosity observed in the optical and UV.  
 As expected from the RIAF ADAF-type flows, we detect also  nonthermal power law emission components with a range of photon indices 1.76-1.87 (5$\times$10$^{31}$ \lumcgs) for BZ Cam and 1.5-1.8 (9$\times$10$^{31}$ \lumcgs) for V592 Cas. This component is expected to be low at low radiative efficiency and  increase with  accretion rate. We find that this is only about (0.1-0.2)\% of the accretion luminosity of the sources and that the power law component increases with accretion rate, as the power law luminosity is twice as high in V592 Cas.  
 In general, thermal X-rays are mainly from the vicinity of the WD originating from a radiatively inefficient hot accretion flow with NEI conditions. The nonthermal emission can be of the Comptonized part of the flow emission, and/or inverse Comptonized soft disk photons (e.g., corona), and/or originating from scatterings off  the structures in the system (e.g., bipolar winds, disk) or perhaps originates in the jet-like outflow itself as in BZ Cam.
 
We recover a complex iron line structure in 6.0-7.0 keV.  We detect the Fe XXV and Fe XXVI ionization emission lines except for the H-like iron from V592 Cas. None of the line centroids are found at their equilibrium values for both sources even using a free-free continuum, but  they are found at slightly lower energies expected from plasma in NEI which is characteristics of advective hot flows. In addition, we note that the He-like iron is a blend of the forbidden, intercombination and resonance lines and the shifts of centroids towards lower energies would be towards the forbidden emission line which justifies the NEI conditions.  Tables 1 and 2 reveal that NEI model fits yield a physically better description of the data with good \rchisq\ values at higher Confidence Level in comparison to the rest of the models.  We suggest that the detected radiatively inefficient advective hot flows portray a weakly collisional plasma in both NLs where magnetic turbulence effects (as a result of magneto-rotational instabilities or alfvenic waves) drive the plasma away from equipartition as the ions and electrons are heated differently. 
  
 The PDSs we calculate for the two NL sources (for BZ Cam at 1$\sigma$ only) show a peaked noise continuum component most likely originating from the hot flow. We find transitions  of the optically thick flow into the inner advective hot flow at around radii  R$_{tr}$ $\sim$ 7.4$\pm$2.4$\times$10$^9$ cm for BZ Cam,  5.5$\pm$2.0$\times$10$^9$ cm for MV Lyr, and $>$3.0$\times$10$^9$ cm for V592 Cas, where we also detect a QPO from this system (V592 Cas). All is in line with expectations from the X-ray binary sources (NS or BH) as their accretion rate increases the noise in the PDS shifts  to  higher frequencies and become more coherent.  We find that the PDS of our NL sources are consistent with the DNe PDSs calculated in other studies in quiescence and in high state (NLs are high state CVs).  All fits show the existence of a 6.4  keV (6.2-6.5 keV)  iron reflection line (with a width of 0.2-0.4 keV) which indicates the existence of a reflection spectrum or rather existence of reflection effects on the spectrum of both of the sources, but we do not detect the Compton reflection hump around 20-30 keV. Modeling the spectrum with available reflection models do not improve the fits (even when the power law components are removed). 
 
Finally, we detect H-like iron absorption line (Fe XXVI) in both of the sources that yield large blueshifts. For BZ Cam we detect a P Cygni-type profile for the H-like iron owing to its  shape in emission and absorption. A simple Doppler
calculation yields outflow velocities of (4500-8700) km s$^{-1}$ and the outflow is directed towards the observer (i.e., jet-like outflow) .  This is the first time a jet-like outflow is detected in the X-rays from an accreting CV.  We suggest that the bow-shock nebula around
BZ Cam may be powered by the fast-collimated outflow from the binary (jet power $\sim$ several $\times$10$^{31}$ \lumcgs; possibly out to $\sim$10$^{34}$ \lumcgs given the accretion luminosity) and the interaction and associated mass can mimic a classical nova remnant.  BZ Cam possibly dissipates most of the gravitational power in the X-ray emitting region via  jet-like outflows as a result of the radiatively inefficient advective accretion detected in the X-rays.
The same H-like iron absorption in V592 Cas has no significant H-like emission line in the same range, thus it is possible that even if there is an outflow in the X-rays, it is not directed towards the observer.  The absorption feature of V592 Cas  yields blueshifts out to  3000 km s$^{-1}$ consistent with the wind velocities in the optical and UV energy bands. 
The detected  Fe XXVI absorption features indicate existence of warm absorber effects in both BZ Cam and V592 Cas as observed in the X-ray binaries. 
These two NL systems set example to characteristics of accretion in high state CVs with bipolar fast wind outflows as observed in the optical and UV (for BZ Cam in the X-rays detected in this study). Their X-ray characteristics shed light on the theory of accretion in high accretion rate white dwarf binaries, as they portray existence of radiatively inefficient advective hot flows with extended geometries and magnetically aided accretion and outflows, revealing a consensus of accretion physics around compact objects.

\begin{acknowledgments}
The authors acknowledge the \nustar\ Observatory for performing the observations
of BZ Cam, V592 Cas.  This work was partially supported by the Vaughan Family fund at UTSA. 
Authors thank Valery Suleimanov  and Tomaso Belloni for comments on the manuscript. SB thanks  E. Sion, P. Szkody, D. Bisikalo and R. Gonzalez-Riestra for helpful discussions
and support with the \nustar\ proposal.  PG is pleased to thank William P. Blair at the Henry A. Rowland Department of Physics and Astronomy, Johns Hopkins University, Baltimore, Maryland, 
for his kind hospitality.

\end{acknowledgments}


\end{document}